\documentclass[journal,comsoc]{IEEEtran}
\usepackage[T1]{fontenc}

\ifCLASSINFOpdf
  \usepackage[pdftex]{graphicx}
  \graphicspath{{./png/}{./pdf/}{./jpeg/}}
  \DeclareGraphicsExtensions{.pdf,.jpeg,.jpg,.png}
\else
  \usepackage[dvips]{graphicx}
  \graphicspath{{./eps/}}
  \DeclareGraphicsExtensions{.eps}
\fi
\usepackage{amsmath}
\usepackage[cmintegrals]{newtxmath}
\ifCLASSOPTIONcompsoc
  \usepackage[caption=false,font=normalsize,labelfont=sf,textfont=sf]{subfig}
\else
  \usepackage[caption=false,font=footnotesize]{subfig}
\fi

\usepackage{stfloats}
\usepackage{todonotes}
\usepackage[detect-weight=true, detect-family=true, per-mode=symbol]{siunitx}
\usepackage{booktabs}
\renewcommand{\vec}[1]{\ensuremath{\boldsymbol{#1}}}

\usepackage[acronym]{glossaries}
\newacronym{4G}{4G}{fourth generation}
\newacronym{ADC}{ADC}{analogue-to-digital converter}
\newacronym{ADS}{ADS}{advanced design system}
\newacronym{AoA}{AoA}{angle of arrival}
\newacronym{AoD}{AoD}{angle of departure}
\newacronym{AP}{AP}{Access Point}
\newacronym{AS}{AS}{angular spread}
\newacronym{AWGN}{AWGN}{Additive White Gaussian Noise}
\newacronym{BBU}{BBU}{Base Band Unit}
\newacronym{BCC}{BCC}{Bristol City Council}
\newacronym{BER}{BER}{bit error rate}
\newacronym{BF}{BF}{beamforming}
\newacronym{BIO}{BIO}{Bristol Is Open}
\newacronym{BS}{BS}{base Station}
\newacronym{BW}{BW}{bandwidth}
\newacronym{CCDF}{CCDF}{complementary cumulative distribution function}
\newacronym{CDF}{CDF}{cumulative distribution function}
\newacronym{CDMA}{CDMA}{Code Division Multiple Access}
\newacronym{CDT}{CDT}{Centre for Doctoral Training}
\newacronym{COTS}{COTS}{Commercial off-the-shelf}
\newacronym{CP}{CP}{Cyclic Prefix}
\newacronym{CPA}{CPA}{coordinated pilot assignment}
\newacronym{CRC}{CRC}{Cyclic Redundancy Check}
\newacronym{CSI}{CSI}{channel state information}
\newacronym{CSIRO}{CSIRO}{Commonwealth Scientific and Industrial Research Organisation}
\newacronym{CSN}{CSN}{Communication Systems \& Networks}
\newacronym{D2D}{D2D}{device-to-device}
\newacronym{DAC}{DAC}{Digital-to-Analogue Converter}
\newacronym{dB}{dB}{decibels}
\newacronym{dBm}{dBm}{decibel-milliwatts}
\newacronym{DL}{DL}{downlink}
\newacronym{DMA}{DMA}{direct memory access}
\newacronym{DRAM}{DRAM}{dynamic random access memory}
\newacronym{DSP}{DSP}{Digital Signal Processing}
\newacronym{EM}{EM}{Electro-Magnetic}
\newacronym{EPSRC}{EPSRC}{Engineering and Physical Sciences Research Council}
\newacronym{ESD}{ESD}{Electro-Static Discharge}
\newacronym{FFT}{FFT}{Fast Fourier Transform}
\newacronym{FDD}{FDD}{frequency division duplex}
\newacronym{FD-MIMO}{FD-MIMO}{Full Dimension \gls{MIMO}}
\newacronym{FIFO}{FIFO}{first-in first-out}
\newacronym{FITRA}{FITRA}{Fast Iterative Truncation Algorithm}
\newacronym{FOC}{FOC}{Frequency Offset Correction}
\newacronym{FPGA}{FPGA}{field-programmable gate array}
\newacronym{FXP}{FXP}{fixed-point}
\newacronym{GPS}{GPS}{global positioning system}
\newacronym{GPSDO}{GPSDO}{\gls{GPS} disciplined oscillator}
\newacronym{GNSS}{GNSS}{Global Navigation Satellite Systems}
\newacronym{H2T}{H2T}{Host-to-target}
\newacronym{HD}{HD}{High-Definition}
\newacronym{HDL}{HDL}{Hardware Description Language}
\newacronym{ICT}{ICT}{Information and Communications Technology}
\newacronym{ID}{ID}{identity}
\newacronym{IFFT}{IFFT}{Inverse Fast Fourier Transform}
\newacronym{iid}{iid}{independent and indentically distributed}
\newacronym{IO}{IO}{Input/Output}
\newacronym{IP}{IP}{Intellectual Property}
\newacronym{ISI}{ISI}{Inter-Symbol Interference}
\newacronym{IUI}{IUI}{inter-user interference}
\newacronym{JSDM}{JSDM}{Joint Spatial Division Multiplexing}
\newacronym{KPH}{KPH}{kilometres per hour}
\newacronym{LDPC}{LDPC}{low-density parity-check}
\newacronym{LED}{LED}{Light Emitting Diode}
\newacronym{LF}{LF}{Loading Factor}
\newacronym{LIDAR}{LIDAR}{Laser Illuminated Detection \& Ranging}
\newacronym{LO}{LO}{local oscillator}
\newacronym{LOS}{LOS}{line-of-sight}
\newacronym{LS}{LS}{least squares}
\newacronym{LSD}{LSD}{Local Search Detection}
\newacronym{LTE}{LTE}{long-term evolution}
\newacronym{LUT}{LUT}{Look-up Table}
\newacronym{MAC}{MAC}{Medium Access Control}
\newacronym{MF}{MF}{matched filtering}
\newacronym{MRT}{MRT}{maximal ratio transmission}
\newacronym{MCS}{MCS}{modulation and coding scheme}
\newacronym{MGS}{MGS}{modified gram schmidt}
\newacronym{MIMO}{MIMO}{multiple-input, multiple-output}
\newacronym{ML}{ML}{Maximum likelihood}
\newacronym{MSE}{MSE}{Mean Square Error}
\newacronym{MMSE}{MMSE}{Minimum Mean Square Error}
\newacronym{MPD}{MPD}{Message Passing Detection}
\newacronym{MRC}{MRC}{maximal ratio combining}
\newacronym{MU}{MU}{multi-user}
\newacronym{MUI}{MUI}{Multi-User Interference}
\newacronym{MUSIC}{MUSIC}{Multiple Signal Classification}
\newacronym{MWC}{MWC}{mobile world congress}
\newacronym{MXI}{MXI}{Multisystem Extension Interface}
\newacronym{NI}{NI}{National Instruments}
\newacronym{NLOS}{NLOS}{non-line-of-sight}
\newacronym{OBR}{OBR}{Out-of-band power ratio}
\newacronym{OCXO}{OCXO}{Oven-controlled Crystal Oscillator}
\newacronym{OFDM}{OFDM}{Orthogonal Frequency Division Multiplexing}
\newacronym{OTA}{OTA}{over-the-air}
\newacronym{PA}{PA}{Power Amplifier}
\newacronym{PAPR}{PAPR}{peak-to-average power ratio}
\newacronym{PAS}{PAS}{Power-Azimuth Spectrum}
\newacronym{PC}{PC}{Power Control}
\newacronym{PCIe}{PCIe}{peripheral component interconnect express}
\newacronym{PDP}{PDP}{Power Delay Profile}
\newacronym{PDSCH}{PDSCH}{Physical Downlink Shared Channel}
\newacronym{PHY}{PHY}{physical layer}
\newacronym{PL}{PL}{Path-Loss}
\newacronym{PLL}{PLL}{Phase-locked Loop}
\newacronym{PMP}{PMP}{precoding, modulation and \gls{PAPR} reduction}
\newacronym{POC}{POC}{proof-of-concept}
\newacronym{P2P}{P2P}{Peer-to-peer}
\newacronym{ppb}{ppb}{parts-per-billion}
\newacronym{PPC}{PPC}{Pilot Power Control}
\newacronym{PSS}{PSS}{primary synchronisation signal}
\newacronym{PTS}{PTS}{Pilot Time Slot}
\newacronym{PUSCH}{PUSCH}{Physical Uplink Shared Channel}
\newacronym{PXIe}{PXIe}{\gls{PCIe} eXtensions for instrumentation}
\newacronym{QAM}{QAM}{quadrature amplitude modulation}
\newacronym{QPSK}{QPSK}{quadrature phase-shift keying}
\newacronym{QRD}{QRD}{QR decomposition}
\newacronym{RAN}{RAN}{Radio Access Network}
\newacronym{RB}{RB}{resource block}
\newacronym{R&D}{R\&D}{Research and development}
\newacronym{RE}{RE}{Resource Element}
\newacronym{RF}{RF}{radio frequency}
\newacronym{RIO}{RIO}{reconfigurable input/output}
\newacronym{RRH}{RRH}{remote radio head}
\newacronym{RSS}{RSS}{Received Signal Strength}
\newacronym{RT}{RT}{Ray-tracing}
\newacronym{RTM}{RTM}{Rear Transition Module}
\newacronym{RX}{RX}{Receive}
\newacronym{SC}{SC}{Single-Carrier}
\newacronym{SDMA}{SDMA}{Space-Divison Multiple Access}
\newacronym{SDR}{SDR}{software-defined radio}
\newacronym{SER}{SER}{Symbol Error Rate}
\newacronym{SINR}{SINR}{signal to interference plus noise ratio}
\newacronym{SIR}{SIR}{Signal to Interference Ratio}
\newacronym{SM}{SM}{Spatial Modulation}
\newacronym{SMA}{SMA}{SubMiniature version A}
\newacronym{SNR}{SNR}{signal to noise ratio}
\newacronym{SPDT}{SPDT}{Single Pole Double Throw}
\newacronym{SISO}{SISO}{Single-Input Single-Output}
\newacronym{SVD}{SVD}{singular value decomposition}
\newacronym{SVS}{SVS}{singular value spread}
\newacronym{T2H}{T2H}{Target-to-host}
\newacronym{TCF}{TCF}{temporal correlation function}
\newacronym{TDD}{TDD}{time division duplex}
\newacronym{TDOA}{TDOA}{Time-Difference-of-Arrival}
\newacronym{TO}{TO}{Timing Offset}
\newacronym{TPC}{TPC}{Transmission Power Control}
\newacronym{TR}{TR}{tone-reservation}
\newacronym{TX}{TX}{Transmit}
\newacronym{TTI}{TTI}{Transmission Time Interval}
\newacronym{U8}{U8}{Unsigned 8-bit}
\newacronym{UDN}{UDN}{ultra-dense network}
\newacronym{UDP}{UDP}{User Datagram Protocol}
\newacronym{UE}{UE}{user equipment}
\newacronym{UI}{UI}{User Interface}
\newacronym{UL}{UL}{uplink}
\newacronym{USRP}{USRP}{universal software radio peripheral}
\newacronym{UoB}{UoB}{University of Bristol}
\newacronym{VLC}{VLC}{VideoLAN Client}
\newacronym{VTC}{VTC}{Vehicular Technology Conference}
\newacronym{WARP}{WARP}{Wireless Open-Access Research Platform}
\newacronym{ZF}{ZF}{zero-forcing}
\begin{document}
\bstctlcite{IEEEexample:BSTcontrol}
%
\title{Performance Characterization of a Real-Time Massive MIMO System with LOS Mobile Channels}
%
%
%

\author{Paul~Harris,~\IEEEmembership{Student Member,~IEEE,}
		Steffen~Malkowsky,~\IEEEmembership{Student Member,~IEEE,}
		Joao~Vieira,
		Erik~Bengtsson,
		Fredrik Tufvesson,~\IEEEmembership{Fellow,~IEEE,}
		Wael~Boukley~Hasan,~\IEEEmembership{Student Member,~IEEE,}
		Liang~Liu,~\IEEEmembership{Member,~IEEE,}
        Mark~Beach,~\IEEEmembership{Member,~IEEE,}
        Simon~Armour,
        and Ove~Edfors,~\IEEEmembership{Member,~IEEE}
\thanks{P. Harris, W.B. Hassan, M.A. Beach and S. Armour are with the \gls{CSN} Group at the University of Bristol, U.K.} \thanks{S. Malkowsky, J. Vieira, E. Bengtsson, F. Tufvesson, L. Liu and O. Edfors are with the Dept. of Electrical and Information Technology, Lund University, Sweden.}
}

\maketitle

\begin{abstract}
The first measured results for massive \gls{MIMO} performance in a \gls{LOS} scenario with moderate mobility are presented, with eight users served in real-time using a 100-antenna \gls{BS} at \SI{3.7}{\giga\hertz}. When such a large number of channels dynamically change, the inherent propagation and processing delay has a critical relationship with the rate of change, as the use of outdated channel information can result in severe detection and precoding inaccuracies. For the \gls{DL} in particular, a \gls{TDD} configuration synonymous with massive \gls{MIMO} deployments could mean only the \gls{UL} is usable in extreme cases. Therefore, it is of great interest to investigate the impact of mobility on massive \gls{MIMO} performance and consider ways to combat the potential limitations. In a mobile scenario with moving cars and pedestrians, the massive \gls{MIMO} channel is sampled across many points in space to build a picture of the overall user orthogonality, and the impact of both azimuth and elevation array configurations are considered. Temporal analysis is also conducted for vehicles moving up to \SI{29}{\kilo\meter\per\hour} and real-time \glspl{BER} for both the \gls{UL} and \gls{DL} without power control are presented. For a 100-antenna system, it is found that the \gls{CSI} update rate requirement may increase by 7 times when compared to an 8-antenna system, whilst the power control update rate could be decreased by at least 5 times relative to a single antenna system.
\end{abstract}

\begin{IEEEkeywords}
Massive MIMO, 5G, Testbed, Field Trial, Mobility
\end{IEEEkeywords}

%
\IEEEpeerreviewmaketitle

\section{Introduction}
\glsresetall
%
%
%
%
\IEEEPARstart{M}{assive} \gls{MIMO} has established itself as a key 5G technology that could drastically enhance the capacity of sub-\SI{6}{\giga\hertz} communications in future wireless networks. By taking the \gls{MU} \gls{MIMO} concept and introducing additional degrees of freedom in the spatial domain, the multiplexing performance and reliability of such systems can be greatly enhanced, allowing many tens of users to be served more effectively in the same time-frequency resource \cite{Marzetta2010}. With a plethora of antennas, the law of large numbers averages out the effects of fast-fading (known as 'channel hardening'), noise and hardware imperfections, and also makes such a system extremely robust to antenna or full \gls{RF} chain failures \cite{6736761} \cite{6891254}.

In addition to theoretical work and results such as those documented in \cite{Marzetta2010}, \cite{6736761} and \cite{Hoydis2013}, various institutions from around the world have been developing large-scale test systems in order to validate theory and test algorithms with real data. Rice University presented a 96-antenna \gls{TDD} system in 2012 based upon the \gls{WARP} platform named ARGOS, reporting some cell capacity predictions made by measuring \gls{SINR} \cite{Shepard2012} \cite{Shepard2013}. Around the same time in Australia, the \gls{FDD} Ngara demonstrator \cite{Suzuki2012a} was reported to have achieved a coded \gls{UL} spectral efficiency of 67.26 bits/s/Hz in a lab environment at 638 MHz using practical low-cost hardware. Aligning more closely with the \gls{LTE} standards, Samsung published their work on \gls{FD-MIMO} in \cite{6810440} and \cite{6525612}. This is a massive \gls{MIMO} prototype that utilises a 2D antenna array form factor for deployment feasibility and cell average throughput gains of up to 3.6 have been indicated in system-level simulations. Also exploiting the elevation dimension more aggressively, ZTE reported a field trail of a \gls{TDD} massive \gls{MIMO} prototype in \cite{Ux2015} where 64 transceiver units served 8 \gls{TDD}-\gls{LTE} commercial handsets located at different floor levels in a high-rise building, achieving a \SI{300}{\mega b\per\second} sum rate in \SI{20}{\mega\hertz} of \gls{BW}. Furthermore, their recent announcements at \gls{MWC} indicate significant progress is being made in pre-5G commercial deployments with China Mobile. Coming from a slightly different perspective than the cellular providers, Facebook have developed a 96-antenna massive \gls{MIMO} \gls{POC} within their Connectivity Lab capable of supporting 24 \gls{MIMO} streams. They aim to conclusively demonstrate that massive \gls{MIMO} could provide long-range rural connectivity from city centres and indicated a lab-based achievement of 71 bits/s/Hz in \cite{Facebook}.

Finally, Lund University and the University of Bristol have developed 100-antenna and 128-antenna real-time testbeds capable of serving 12 wireless devices in the same time-frequency resource with a \gls{TDD} \gls{LTE}-like physical layer. The development of these systems underpins the work presented here. Lund University first described their 100-antenna testbed in \cite{Vieira2014a} along with a distributed \gls{FPGA} processing architecture and were the first to present such a system with real-time operation. The University of Bristol through the joint venture \gls{BIO} \cite{BIO} constructed their system the following year and worked jointly with Lund University and \gls{NI} to implement and test a new, centralised \gls{MIMO} processing architecture, which allowed for greater flexibility in ongoing development. Further details on the new architecture can be found in \cite{Malkowsky2016b}. In 2016, two indoor trials were conducted within an atrium at the University of Bristol, and it was shown that spectral efficiencies of 79.4 bits/s/Hz and subsequently 145.6 bits/s/Hz could be achieved whilst serving up to 22 user clients \cite{PaulGlobecom} \cite{PaulSIPS}. These spectral efficiency results are currently world records and indicate the potential massive \gls{MIMO} has as a technology.

However, wireless technology is usually applied in scenarios that will involve some form of mobility, and it is therefore of great interest to investigate the evolution of massive \gls{MIMO} under more dynamic channel conditions. The aforementioned measurement trials have not yet considered the progression of a composite massive \gls{MIMO} channel with mobility, but measured static terminals. In \cite{7510708}, the authors discuss some of the potential issues with channel ageing in a macro-cell massive \gls{MIMO} deployment and propose a novel \gls{UDN} approach for comparison. From simulation results, they illustrate that massive \gls{MIMO} performance could significantly worsen with mobile speeds of just \SI{10}{\kilo\meter\per\hour}, and that the sensitivity of \gls{ZF} to \gls{CSI} errors could make \gls{MF} the more viable option. In \cite{7519076} and \cite{7473866}, the theoretical impact of channel ageing on the \gls{UL} and \gls{DL} performance of massive \gls{MIMO} is evaluated. Interestingly, the analysis shows that a large number of antennas is to be preferred for maximum performance, even under time-varying conditions, and that Doppler effects dominate over phase noise.

In this paper, 8 \glspl{UE} are served by a 100-antenna \gls{BS} in real-time and the first measured results for massive \gls{MIMO} under moderate mobility in \gls{LOS} are presented. With a large set of spatial samples taken across a mobile scenario, the \gls{SVS} is analysed and insight is provided on the impact of azimuth and elevation dominant array configurations. Furthermore, by measuring the full \gls{MIMO} channel at a time resolution of \SI{5}{\milli\second}, temporal analysis in the form of time correlation, \gls{IUI} and channel hardening is conducted for speeds up to \SI{29}{\kilo\meter\per\hour}. Finally, real-time \glspl{BER} measured for the \gls{UL} and \gls{DL} with no power control are presented to provide an indication of the raw performance achieved by the system.

The outline of the remainder of the paper is as follows. Sec.~\ref{System} provides an overview of the testbed and the additional functionality that was implemented to enable meaningful mobility measurements to be made. One measurement scenario is then outlined in Sec.~\ref{Scen} for both a static and mobile case; \gls{SVS}, temporal and \gls{BER} results are presented in Sec.~\ref{Res}; and the paper concludes with Sec.~\ref{conc}.
\section{System Description}
\label{System}
\begin{figure}[!t]
	\centering
	\includegraphics[width=0.75\columnwidth]{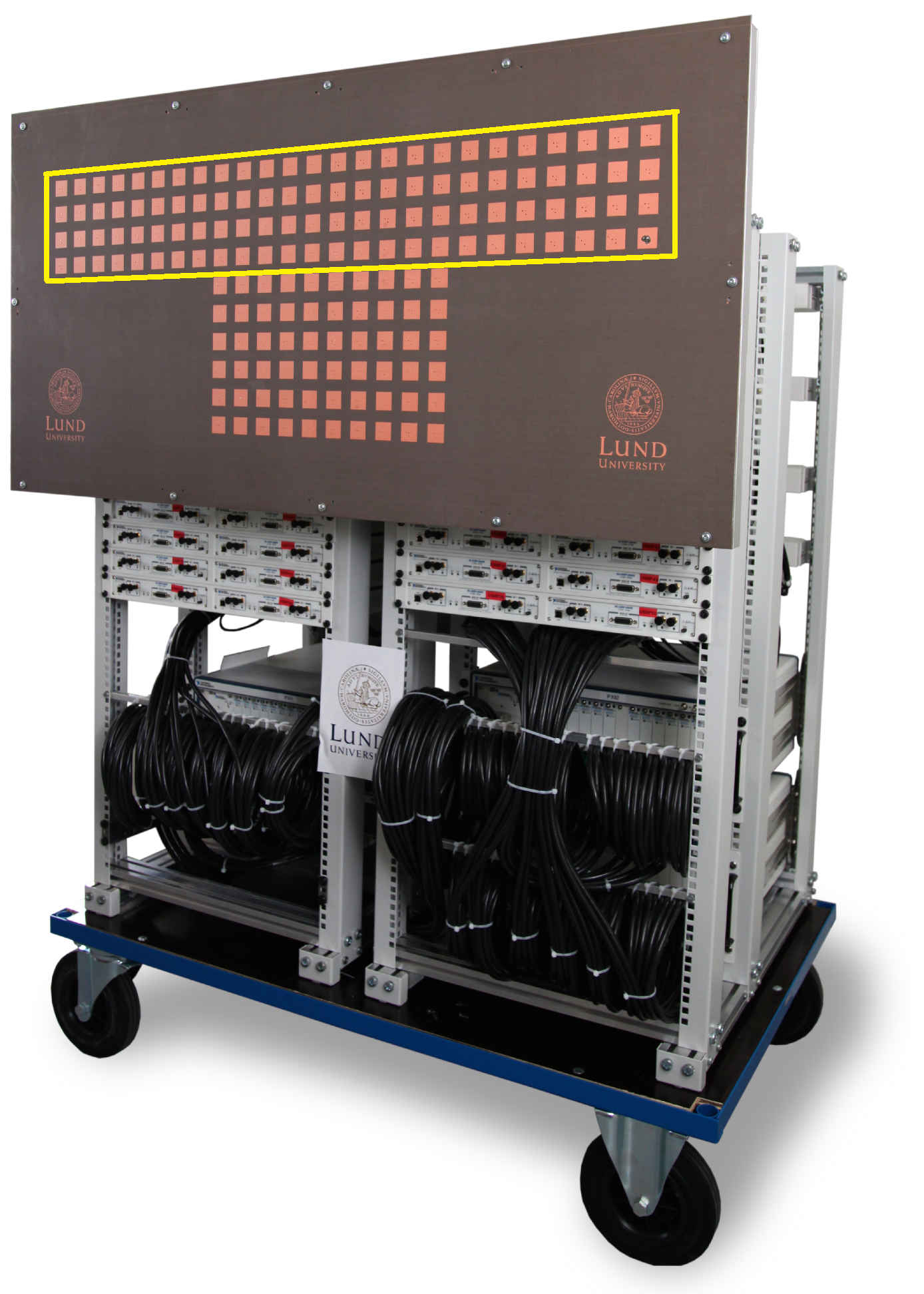}
	\caption{Lund Massive MIMO Basestation with 4x25 array configuration highlighted}
	\label{fig_BS}
\end{figure}

The Lund University massive \gls{MIMO} \gls{BS} pictured in ~\figurename~\ref{fig_BS} consists of 50 \gls{NI} \glspl{USRP}, which are dual-channel \glspl{SDR} with reconfigurable \glspl{FPGA} connected to the \gls{RF} front ends \cite{USRP}. Collectively, these provide 100 \gls{RF} chains, with a further 6 \glspl{USRP} acting as 12 single-antenna \glspl{UE}. It runs with a \gls{TDD} LTE-like \gls{PHY} and the key system parameters can be seen in Table \ref{tab:system_param}. Using the same \gls{PXIe} platform that powers the 128-antenna system, all the \glspl{RRH} and \gls{MIMO} \gls{FPGA} co-processors are linked together by a dense network of gen 3 \gls{PCIe} fabric,
and all software and \gls{FPGA} behaviour is programmed via LabVIEW. A description of the Lund University system along with a general discussion of massive \gls{MIMO} implementation issues can be found in \cite{Malkowsky2016b}. 
The reciprocity calibration approach designed at Lund University and applied in these experiments is detailed in \cite{VieiraREMLT16}. Further detail about the current system architecture and the implementation of a wide data-path \gls{MMSE} encoder/decoder can be found in \cite{PaulGlobecom}, \cite{PaulSIPS} and \cite{7780107}.
\begin{table}
	\renewcommand{\arraystretch}{1.3}
	\caption{System Parameters}
	\centering
	\noindent\begin{tabular}{ll}
		\toprule
		\textbf{Parameter} & \textbf{Value}  \\
		\midrule
		\# of BS Antennas & 100  \\
		\# of UEs & 12 \\
		Carrier Frequency & 1.2-\SI{6}{\giga\hertz} (\SI{3.7}{\giga\hertz} used) \\
		Bandwidth		  & \SI{20}{\mega\hertz}  \\
		Sampling Frequency  & \SI{30.72}{\mega S\per\second} \\
		Subcarrier Spacing  & \SI{15}{\kilo\hertz} \\
		\# of Subcarriers  &  2048 \\
		\# of Occupied Subcarriers  & 1200 \\
		Frame duration  & \SI{10}{\milli\second} \\
		Subframe duration & \SI{1}{\milli\second} \\
		Slot duration & \SI{0.5}{\milli\second} \\
		TDD periodicity &   1 slot  \\	
		\bottomrule
	\end{tabular}
	\label{tab:system_param}
\end{table}
\subsection{Antenna Array}
The Lund antenna array in~\figurename~\ref{fig_BS} consists of half-wavelength spaced 3.7 GHz patch-antennas each with horizontal and vertical polarization options. For the results reported here, the 4x25 azimuth dominated portion highlighted in the image was used, with alternating horizontal and vertical polarizations across the array.

\subsection{Channel Acquisition}
The system is defined as having $M$ antennas, $K$ users and $N$ frequency domain resource blocks of size $R_{b}$. For this paper, we specifically define resource block to refer to a grouping of 12 \gls{OFDM} subcarriers, with each 1200 subcarrier \gls{OFDM} symbol consisting of 100 resource blocks.
The estimate of the \gls{UL} for each 12 subcarrier resource block $r$ is found as
\begin{equation}
\vec{H}_{r} = \vec{Y}_{r}\vec{P}^{*}_{r}
\end{equation}
where $\vec{H}_{r}$ is the $M \times K$ channel matrix, $\vec{Y}_{r}$ is the $M \times R_{b}$ receive matrix and $\vec{P}^{*}_{r}$ is the diagonal $R_{b} \times K$ conjugate uplink pilot matrix. Each \gls{UE} sends an \gls{UL} pilot for each resource block on a subcarrier orthogonal to all other users and the \gls{BS} performs least-square channel estimation. Since all pilots have a power of 1, $\vec{Y}_{r}\vec{P}^{*}_{r}$ can be used rather than $\vec{Y}_{r} / \vec{P}^{*}_{r}$.
All \gls{MIMO} processing in the system is distributed across 4 Kintex 7 \gls{FPGA} co-processors, each processing 300 of the 1200 subcarriers. In order to study the channel dynamics under increased levels of mobility, a high time resolution is desired during the capture process, which equates to a high rate of data to write to disk. To address this, a streaming process was implemented that uses the on-board \gls{DRAM} to buffer the raw \gls{UL} subcarriers.
$\vec{Y}_{r}$ is recorded in real-time to the \gls{DRAM} at the measurement rate, $T_\text{meas}=\SI{5}{\milli\second}$, and this data is then siphoned off to disk at a slower rate. 
Using this process with \SI{2}{\giga B} of \gls{DRAM} per \gls{MIMO} processor, we were able to capture the full composite channel for all resource blocks a \SI{5}{\milli\second} resolution for \SI{65}{\second}.
\\

The channel was sampled at least once every half-wavelength distance in space to give an accurate representation of the environment. The maximum permissible speed of mobility, $v_\text{max}$, is thus given by
\begin{equation}
v_\text{max} = \frac{\lambda}{2T_\text{meas}}
\label{speed}
\end{equation}
This results in a maximum speed of \SI{8.1}{\meter\per\second} or approximately \SI{29}{\kilo\meter\per\hour} for temporal analysis of the channel data. For the \gls{SVS} results, this constraint does not apply as the aim is to simply provide a large spatial sample set. It should also be noted that the testbed estimates the full channel every \SI{0.5}{\milli\second} (1 time slot) for real-time operation, so the speed limitation is also not applicable for the real-time \glspl{BER}.

\subsection{Post-Processing}
\subsubsection{SVS}
\label{Post}
The \gls{SVS} is one of the most powerful ways to evaluate the joint orthogonality of the user channel vectors in a \gls{MIMO} system \cite{1192168}. Our $M\times K$ channel matrix for one resource block $r$ can be described in terms of its \gls{SVD} \cite{paulraj2003introduction} as

\begin{equation}
\vec{H}_{r}=\vec{U}_{r}\vec{\Sigma}_{r}\vec{V}_{r}^{H}
\end{equation}
where $\vec{U}_{r}$ and $\vec{V}_{r}$ represent the left and right unitary matrices, and $\vec{\Sigma}_{r}$ is the $M\times K$ diagonal matrix containing the singular values $\sigma_{1,r},\sigma_{2,r},...,\sigma_{K,r}$ sorted in decreasing order. The \gls{SVS} is then defined as

\begin{equation}
\kappa_{r}=\frac{\sigma_{1,r}}{\sigma_{K,r}},	
\end{equation}
i.e. the ratio of the largest to the smallest singular value. A large \gls{SVS} indicates that at least two of the user column vectors are close to parallel and spatially separating these users will therefore be difficult, whereas a $\kappa_{r}$ of one (0 \gls{dB}) represents the ideal case where all the user channel vectors are pairwise orthogonal. 

In order to obtain accurate results for the \gls{SVS} that solely represent the achievable spatial separation, it is important to remove path-loss differences among \glspl{UE} by applying a form of normalization to the raw captured matrix, which we denote as $\vec{H}^{\text{raw}}_{r}$. To achieve this, the first normalization described in \cite{7062910} was applied. This normalization
ensures that the average energy across all $N$ resource blocks and $M$ antennas for a given \gls{UE} in $\vec{H}^{\text{raw}}_{r}$, denoted as $\vec{h}^{\text{raw}}_{i,r}$ for user $i$, is equal to one. This is achieved through

\begin{equation}
\vec{h}_{i,r}^{\text{norm}} = \sqrt{\frac{MN}{\sum\limits_{r=1}^{N}\|\vec{h}_{i,r}^{\text{raw}}\|^2}}\vec{h}_{i,r}^{\text{raw}}
\end{equation}
where $\vec{h}_{i,r}^{\text{norm}}$ is the $i$th column of the normalised channel matrix $\vec{H}^{\text{norm}}_{r}$. This can also be thought of as applying perfect power control.
\\
\subsubsection{Temporal Analysis}
To evaluate the change in the multi-antenna channels under mobility, an analog of the \gls{TCF} \cite{molisch2010wireless} was calculated. By introducing a time dependence on the measured channels, i.e., the channel vector corresponding to user $i$ at resource block $r$ and at time $t$ is denoted by $\vec{h}^{\text{raw}}_{i,r}[t]$, the \gls{TCF} was defined as
\begin{equation}
\text{TCF}_{i}(\tau) = \frac{\mathbf{E}\{ | \vec{h}^{\text{raw}}_{i,r}[t-\tau]^H \; \vec{h}^{\text{raw}}_{i,r}[t] | \} }{ \mathbf{E}\{ \vec{h}^{\text{raw}}_{i,r}[t]^H  \; \vec{h}^{\text{raw}}_{i,r}[t] \}},
\end{equation}
where $\mathbf{E}\{ \}$ denotes the expectation operator. At a given time lag $\tau$, the expectation is computed according to its definition, but also by averaging over all resource blocks for better statistics.

The instantaneous \gls{IUI} between two users was also evaluated, i.e. users $i$ and $u$ with $u \neq i$ by the normalized expected inner product
\begin{equation}
\text{Int}_{i,u}^r [t] = \frac{|\vec{h}^{\text{raw}}_{i,r}[t]^H  \vec{h}^{\text{raw}}_{u,r}[t]|}  {\sqrt{\vec{h}^{\text{raw}}_{i,r}[t]^H \vec{h}^{\text{raw}}_{i,r}[t] \; \vec{h}^{\text{raw}}_{u,r}[t]^H \vec{h}^{\text{raw}}_{u,r}[t]}}
\end{equation}
This provides a metric to evaluate to what extent the, so-called, massive \gls{MIMO} favorable propagation conditions \cite{BjornsonLM15} hold in a practical system.

\subsection{User Equipment}
\begin{figure}[!t]
	\centering
	\includegraphics[width=\columnwidth]{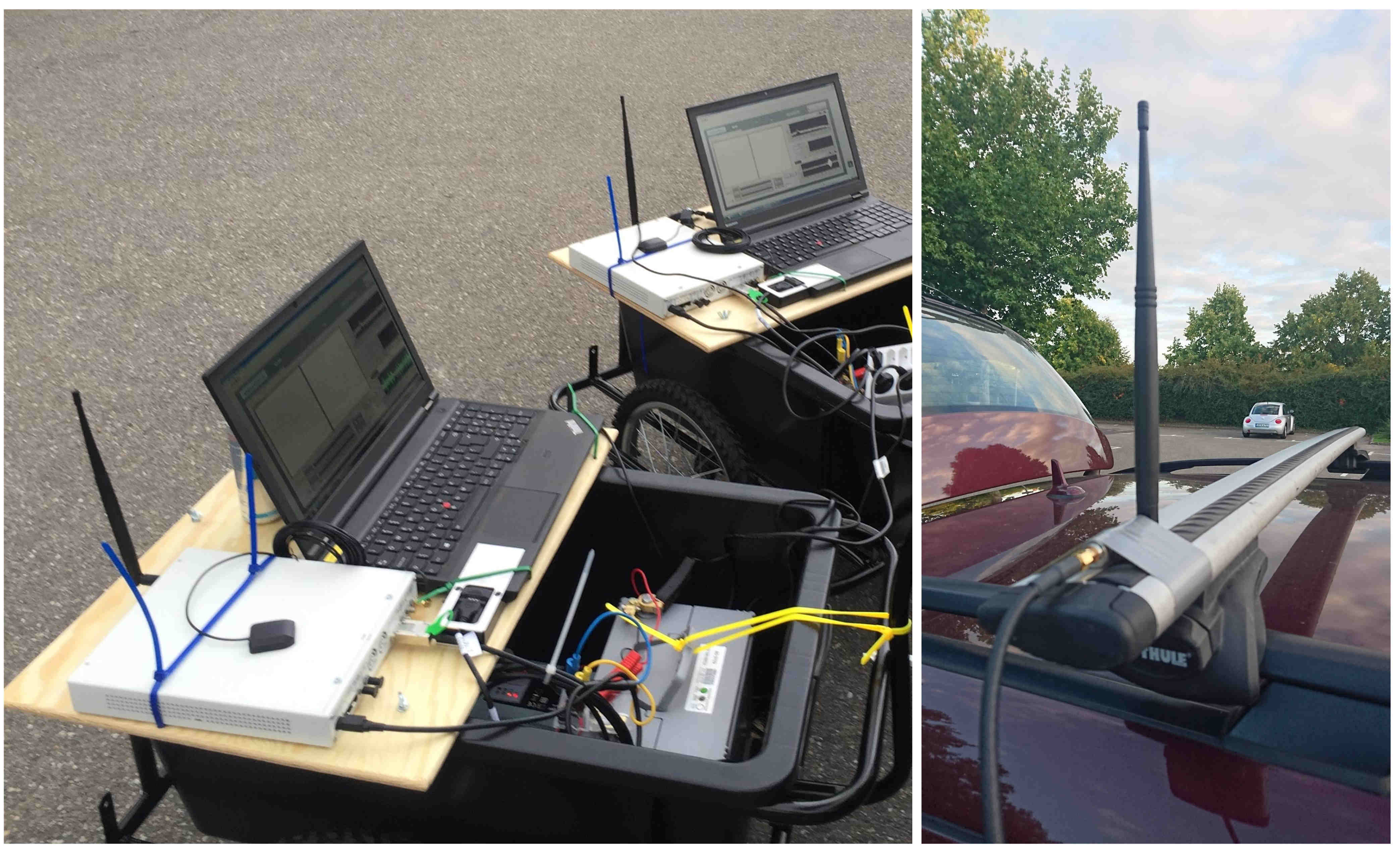}
	\caption{User Equipments. Left: pedestrian carts. Right: car mounting.}
	\label{fig_UEs}
\end{figure}

As mentioned at the beginning of Sec.~\ref{System}, each \gls{UE} is a two-channel \gls{USRP}, four of which were used in these measurements to provide a total of 8 spatial streams.
The \glspl{USRP} were mounted in carts to emulate pedestrian behaviour and in cars for higher levels of mobility, as shown in ~\figurename~\ref{fig_UEs}.
Sleeve dipole antennas where used in each case. On the carts, the dipole antennas were attached directly to both \gls{RF} chains of the \gls{USRP} with vertical polarisation. With a fixed spacing of 2.6$\lambda$, these can either be considered as two devices in very close proximity or as a single, dual-antenna device. In~\figurename~\ref{fig_UEs}, the \glspl{USRP} shown in carts each have only one antenna connected.
For the cars, the antennas were roof mounted with vertical polarization on either side of the car, giving a spacing of approximately \SI{1.7}{\meter}. Each \gls{UE}'s
\gls{RF} chain operates with a different set of frequency-orthogonal pilots and synchronises \gls{OTA} with the \gls{BS} using the \gls{PSS} broadcast at the start of each \SI{10}{\milli\second} frame. The \gls{PSS} was transmitted using a static beam pattern and the \glspl{LO} of the \glspl{UE} were \gls{GPS} disciplined to lower carrier frequency offset to improve sync stability.

\section{Measurement Scenario}
\label{Scen}
In this section, the measurement scenario and two configurations used to obtain both static and mobile measurements are described. Both measurements can be considered as predominantly \gls{LOS} with an expected channel estimation \gls{SNR} range of approximately \SI{20}{\decibel} to \SI{30}{\decibel} per \gls{RF} chain over the measurement runs considered.

\begin{figure}[!t]
	\centering
	\includegraphics[width=\columnwidth]{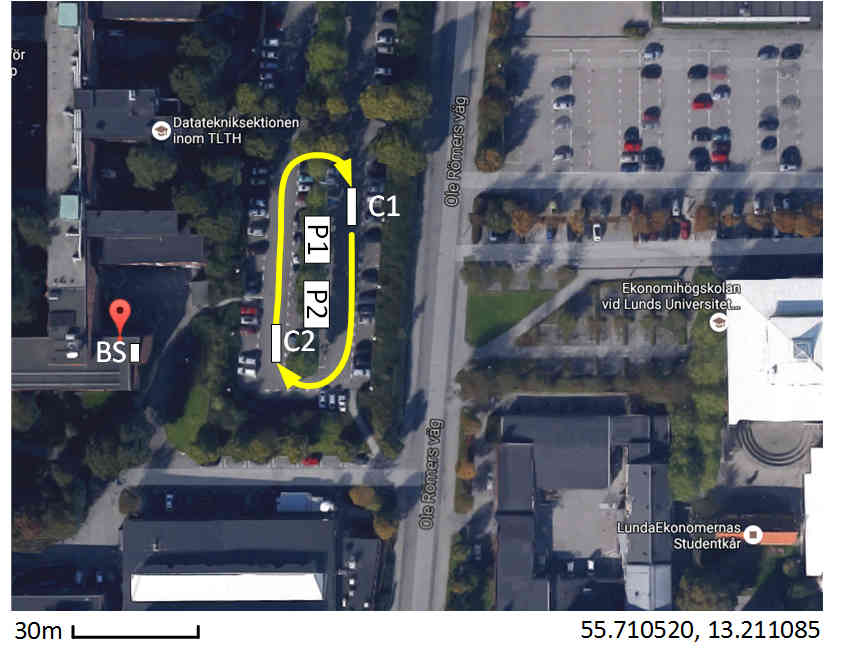}
	\caption{Overview of the measurement scenario at the campus of the Faculty of Engineering (LTH), Lund University, Sweden. Arrows indicate the direction of movement for the cars C1 and C2. Pedestrians P1 and P2 moved within the zones indicated by their white boxes.}
	\label{fig_mob_above}
\end{figure}

\subsection{\gls{LOS} Static Configuration}
To obtain a point of reference for the mobile configuration, a static trial was conducted first using the pedestrian carts. Four dual-antenna \glspl{USRP} acting as 8 \glspl{UE} were placed \SI{32}{\meter} away from the \gls{BS} in a parallel line where car 2 (C2) is shown in ~\figurename~\ref{fig_mob_above}. The carts were well separated (approximately \SI{2}{\meter} apart), each \gls{UE} transmitted with a fixed power of approximately \SI{0}{\decibel m}, and $\vec{Y}_{r}\forall \ r$ was recorded for \SI{60}{\second}. Movement was kept to a minimum within the surrounding environment during capture period.

\subsection{\gls{LOS} Mobile Configuration}
For the mobile trial, a mixture of both pedestrian and vehicular \glspl{UE} were introduced, so that it would be possible to observe how the massive \gls{MIMO} channel behaves over time in a more dynamic situation. Each user again transmitted with the same fixed power level and $\vec{Y}_{r}\forall \ r$ was captured for a \SI{30}{\second} period. 
Two pedestrian carts, indicated by P1 and P2, moved pseudo-randomly at walking pace to and from one another for the measurement duration, whilst two cars, shown as C1 and C2, followed the circular route shown. For the temporal results concerning the cars, it was ensured that the captures analysed were from a period of the scenario where the cars did not exceed our maximum $\lambda / 2$ measurement speed of \SI{29}{\kilo\meter\per\hour}. Over the course of the entire capture, the cars completed approximately two laps and arrived back at the starting position indicated in ~\figurename~\ref{fig_mob_above}. With the cars moving in this pattern, the devices are, on average, more distributed in the azimuth, but when C1 and C2 pass in parallel to the pedestrian carts they become more clustered in a perpendicular line to the \gls{BS}.

\section{Results}
\label{Res}
Results from the experiments are shown here in three stages. The \gls{SVS} results are inspected first, considering the impact of azimuth and elevation dimension reductions on the spatial orthogonality. These results illustrate the range of orthogonality experienced over all sampled points in space as the devices were moved. An example of channel hardening is then shown second, along with results for correlation and \gls{IUI} over time between a car \gls{UE} and a pedestrian \gls{UE}, providing some insight into the temporal nature of the massive \gls{MIMO} channels. Finally, the uncoded \gls{UL} and \gls{DL} real-time \gls{BER} performance is presented.
\begin{figure}[!t]
	\centering
	\includegraphics[width=\columnwidth]{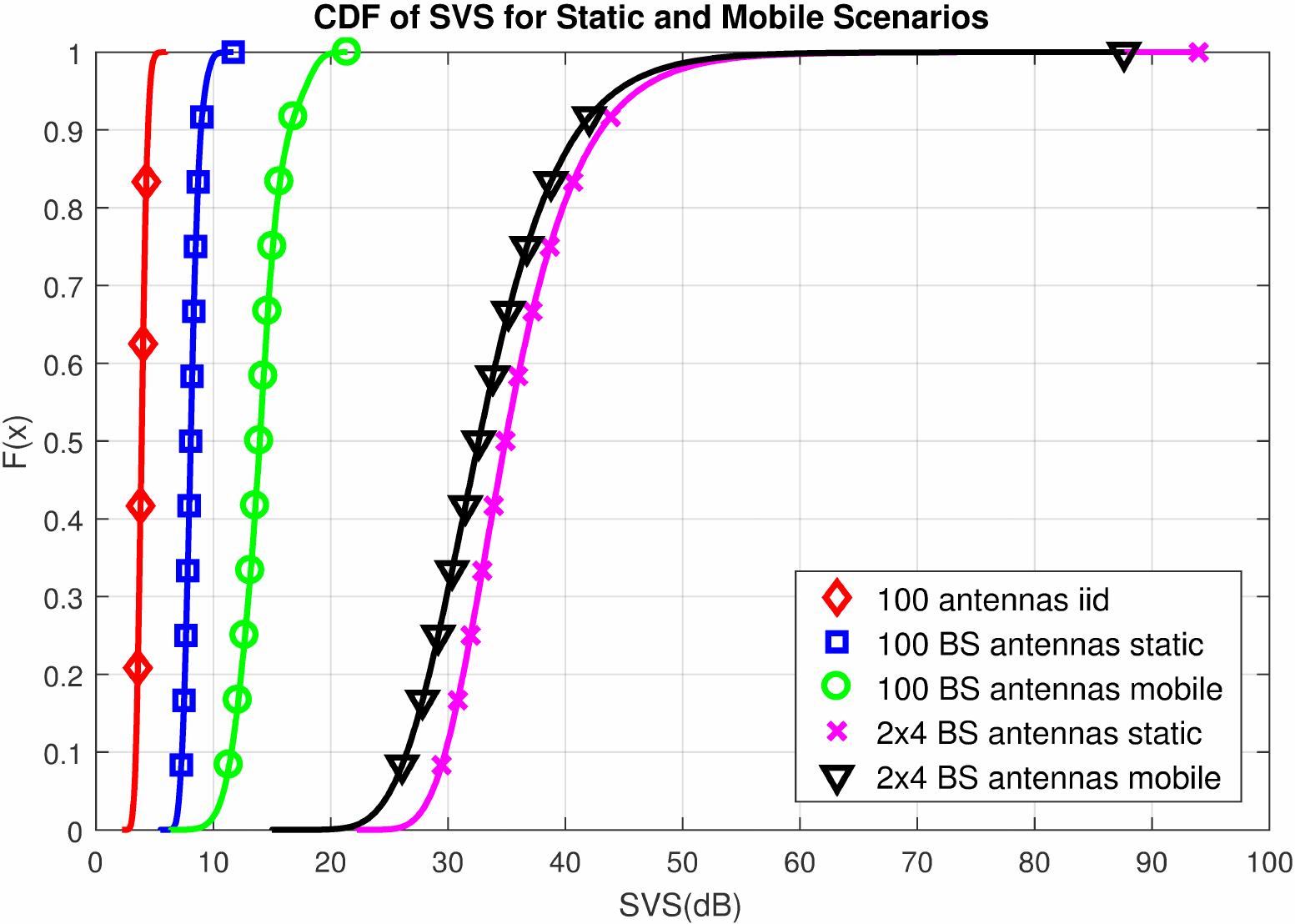}
	\caption{CDF of SVS for static and mobile scenarios using 100 and 8 antennas at the \gls{BS}.}
	\label{fig_SVS_stat_vs_mob}
\end{figure}
\subsection{Singular Value Spread}
As described in Sec.~\ref{Post}, the \gls{SVS} is an effective way to study the pairwise orthogonality of the user channel vectors and thus the achievable spatial multiplexing performance for a \gls{MIMO} system. The \gls{SVS} results are presented here as empirical \gls{CDF} plots for all $N$ resource blocks and captures, i.e. across time and frequency. The result is 600,000 data points for the \SI{30}{\second} mobile scenario and 1,200,000 for the \SI{60}{\second} static scenario. The static and mobile scenario are compared first to see how much the \gls{CDF} spreads out when spatial samples across the entire measurement period are considered. These results are shown in~\figurename~\ref{fig_SVS_stat_vs_mob} for the full 100 antenna case ($4\times 25$) and an 8 antenna ($2\times 4$)\footnote{$2\times4$ was chosen to provide a similar array shape and azimuth dominance to the full $4\times25$, 100 antenna case.} case, providing the minimum for standard \gls{MU}-\gls{MIMO} with 8 \glspl{UE}. The theoretical \gls{iid} case for a $100\times8$ system is also shown as a reference.
The median value for the $100\times8$ \gls{iid} case here is a little over 3.8 \gls{dB} with barely any variance on the \gls{CDF} curve. It can be seen that the static reference measurement with 100 antennas has its median shifted from this by \SI{4.2}{\decibel} up to \SI{8}{\decibel}, but it also has a similar stability to the \gls{iid} case with very small levels of variance. This indicates a very good level of spatial separation in \gls{LOS} that is likely to be predominantly limited by the angular resolution provided by the azimuth dimension of the \gls{BS} array (25 antenna elements) and the dual-\gls{UE} terminals with only 2.6$\lambda$ spacing between antennas. The result may have improved if 8 single-antenna devices were served with a more reasonable spacing. With only 8 antennas at the \gls{BS}, as would be the case in a standard \gls{MU}-\gls{MIMO} system, the median value in the static case is approximately \SI{35}{\decibel} with an upper tail reaching out towards \SI{94}{\decibel}. This indicates that the user vectors are highly parallel and it will be far more difficult to establish reliable spatial modes.
\begin{figure}[!t]
	\centering
	\includegraphics[width=\columnwidth]{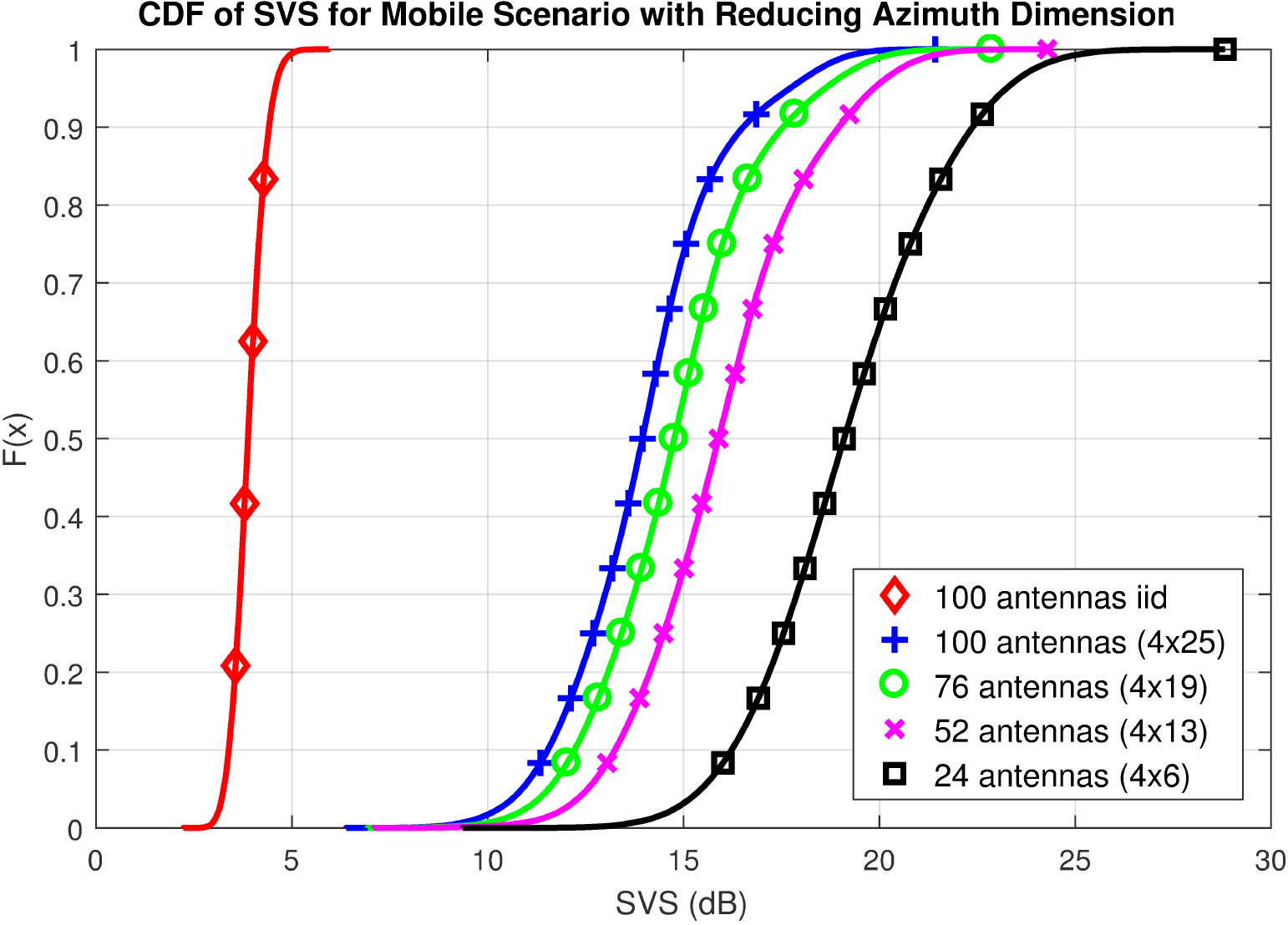}
	\caption{CDF of SVS for mobile scenario with reducing azimuth dimension.}
	\label{fig_SVS_az}
\end{figure}
In the mobile case with 100 \gls{BS} antennas the median shifts to nearly \SI{14}{\decibel}, twice the magnitude of the static case. The best case scenario at the lower tail of the curve comes in line with the static measurement, but the variance over the measurement period has increased, reaching a peak \gls{SVS} of \SI{21}{\decibel}. However, the upper tail of the curve is still relatively small with a 90th percentile of \SI{16.5}{\decibel}, indicating a good level of stability and a restriction in the extremity of the variations. With only 8 \gls{BS} antennas, the median \gls{SVS} increases to \SI{32.5}{\decibel}, and the majority of the \gls{CDF} plot up to the \SI{41}{\decibel} 90th percentile has decreased \SI{2.5}{\decibel} from the static case. This illustrates that with only 8 antennas, separating the static users equally spaced by approximately \SI{2}{\meter} in a single parallel line proved more difficult than the dispersed mobility scenario. In summary, these results indicate that the variation between spatial channel magnitudes can be kept below \SI{16.5}{\decibel} for 90\% of the \gls{LOS} scenario shown with an $M$ to $K$ ratio of 12.5.
\subsection{Azimuth vs. Elevation}
One topic of considerable interest for massive \gls{MIMO} deployments is an optimal array configuration. Massive \gls{MIMO} performance is generally expected to be higher in a cellular scenario when using a more azimuth dominant configuration due to a higher spread in the angular arrival of multipath components, but the extent to which this affects actual performance in real scenarios will help determine the compromises which can be made for a feasible deployment. In addition, when the scenario is \gls{LOS} or more Rician in nature, the dominant components will become directional beams and higher performance would be expected when the array dimension is largest in the plane the \glspl{UE} are spread. For example, when the \glspl{UE} are placed in a line perpendicular to the \gls{BS} with differing distances, one would typically prefer additional resolution in the elevation to resolve them. ~\figurename~\ref{fig_SVS_az} shows the \gls{SVS} for the 30 second mobile scenario as the azimuth dimension of the array is reduced in intervals of one quarter, starting from the outside edges and moving in\footnote{It was only possible to reduce by multiples of 2 due to the 4x25 array configuration.}. Moving through the first 3 configurations, a clear shift of the curve by steps of approximately \SI{1}{\decibel} can be seen and only a slight growth in the upper tail. Once the $4\times6$ case is reached, the median \gls{SVS} magnitude has increased by \SI{5}{\decibel} to \SI{19}{\decibel}.
\begin{figure}[!t]
	\centering
	\includegraphics[width=\columnwidth]{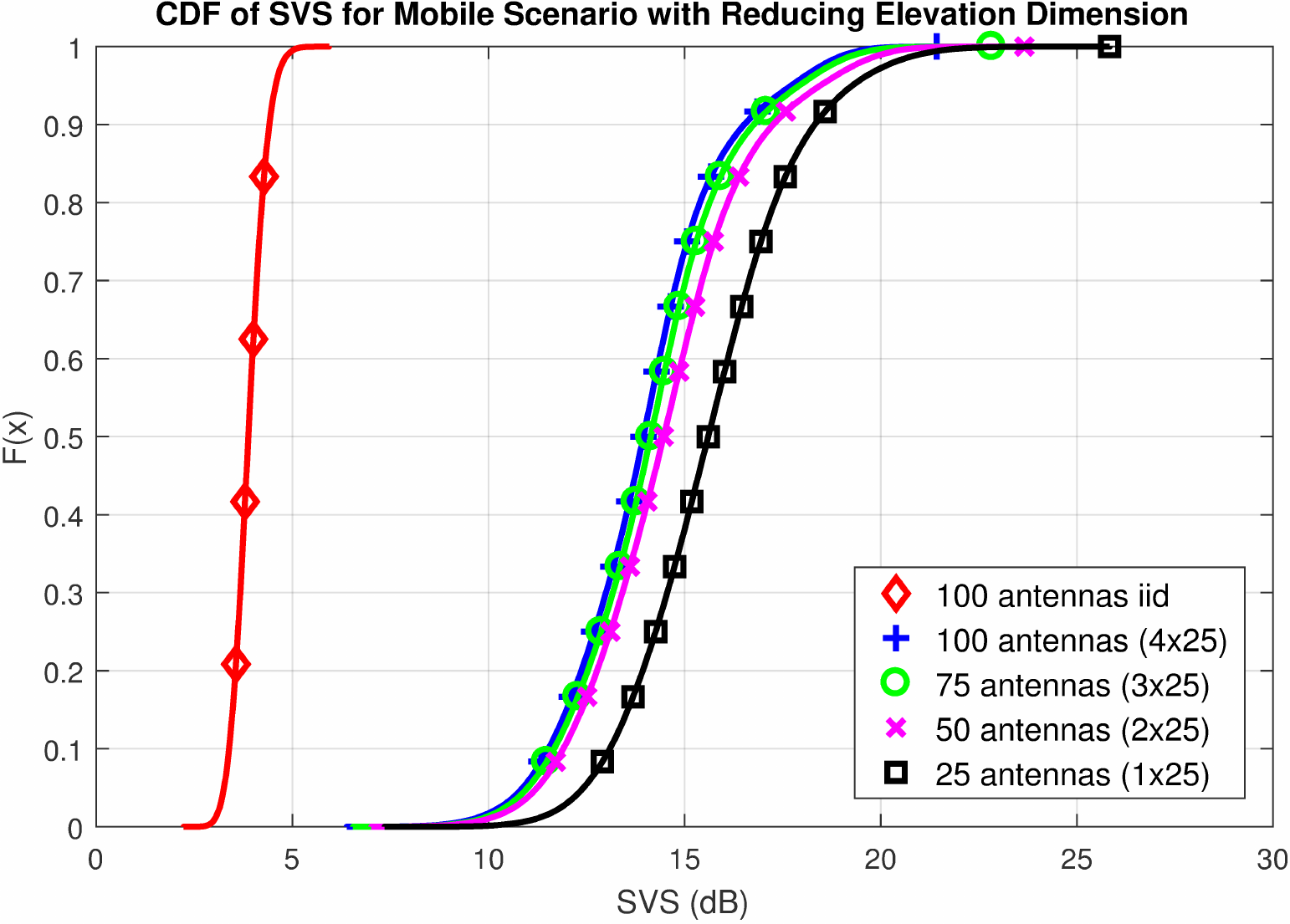}
	\caption{CDF of SVS for mobile scenario with reducing elevation dimension.}
	\label{fig_SVS_elev}
\end{figure}
In ~\figurename~\ref{fig_SVS_elev} the \gls{SVS} results for the same mobile scenario are shown, but this time with the elevation dimension of the array reducing by 1 row (one quarter) at a time. It is immediately apparent that the results for the first three steps do not differ significantly at all from the full 100 antenna case. The upper tail has extended by just over \SI{1}{\decibel}, but the median and 90th percentile have increased by only \SI{0.2}{\decibel} and \SI{0.5}{\decibel}, respectively. As the antenna dimensions are reduced, it can be seen that all configurations are closer to the full 100 antenna curve than in the reduced azimuth cases. Even the $1\times25$ case where there is no elevation resolution outperforms the $4\times13$ case.
This shows that for this particular scenario, there is more to gain from the azimuth dimension, and a similar level of performance could be obtained by halving the number of antennas in the elevation dimension. However, looking at the two \gls{CDF} plots, one could argue that the difference is not significant enough to prefer the extended azimuth arrangement, particularly with the deployment difficulties it could introduce for mast mounting. This could be indicative that a more symmetrical arrangement such as that shown for \gls{FD-MIMO} in \cite{6810440} and \cite{6525612} would provide sufficient performance when compared to an azimuth dominated configuration. More experiments need to be conducted in urban environments with richer scattering for both \gls{LOS} and \gls{NLOS} situations to confirm this.
\begin{figure}[!t]
	\centering
	\subfloat[]{\includegraphics[width=\columnwidth]{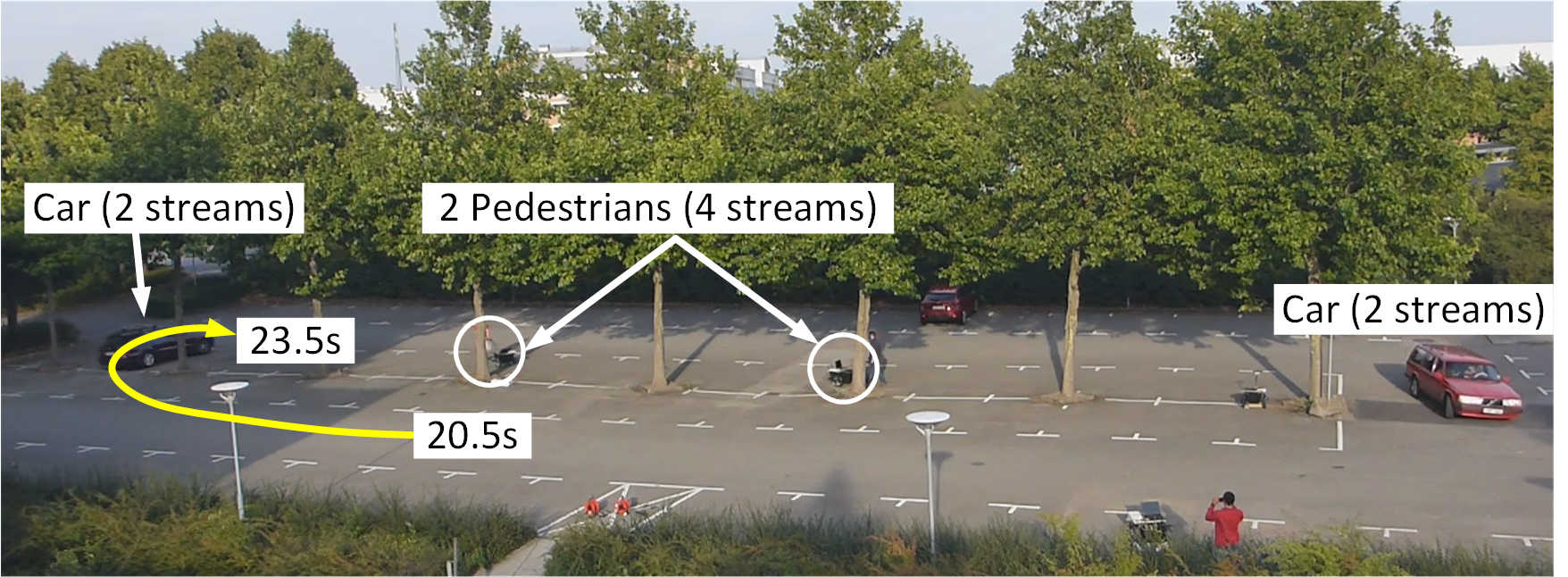}}\\
	\subfloat[]{\includegraphics[width=\columnwidth]{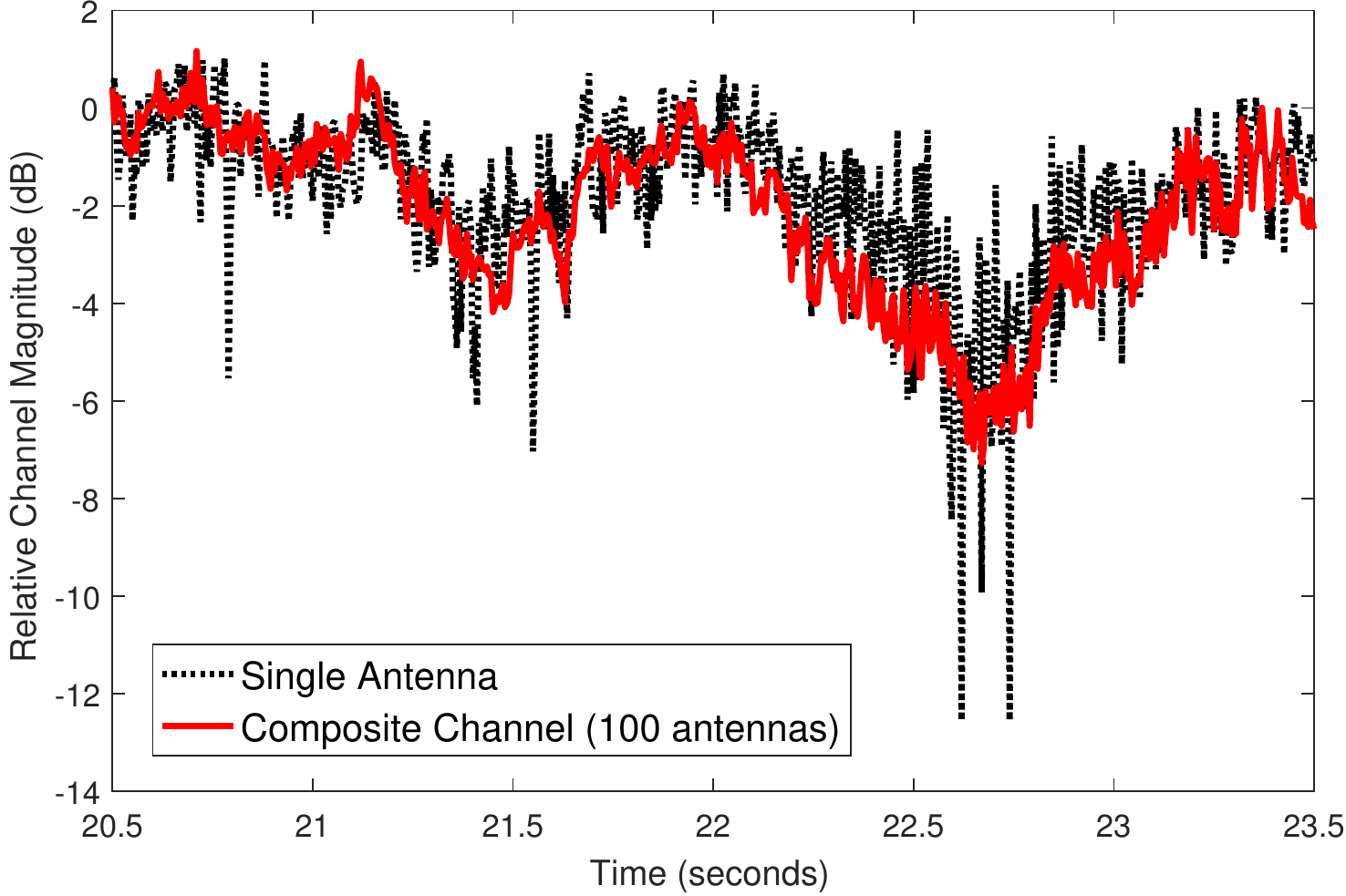}}
	\caption{Resilience to fading. a) View from \gls{BS} with 3 second car-based user path indicated. b) Relative channel magnitude for both a single antenna and the composite \gls{MIMO} channel to the depicted user over the 3 second period.}
	\label{fig_fading}
\end{figure}
\subsection{Temporal Analysis}
The temporal results were based on two different time periods within the 30 second mobility scenario. Channel hardening results are presented first for one car user using a 3 second period of the full 30 second capture, specifically between \SI{20.5}{\second} and \SI{23.5}{\second}. Time correlation and \gls{IUI} results are then shown for a 4 second period between \SI{10}{\second} and \SI{14}{\second} where both cars travel in parallel to the \gls{BS}. For both of these time periods, the vehicle speed remains below \SI{29}{\kilo\meter\per\hour}.
\\
\subsubsection{Channel Hardening}
From theory, fast-fading is shown to disappear when letting the number of \gls{BS} antennas go to infinity, as discussed in \cite{Marzetta2010} and \cite{6736761}.
Whilst the measured scenarios will have been more of a Rician than Rayleigh nature due to the \glspl{UE} being predominantly in \gls{LOS}, it was still possible to inspect the less severe fading dips of a single channel for a single \gls{UE} and evaluate them against the composite channel formed by the $100\times8$ massive \gls{MIMO} system. 
In \figurename~\ref{fig_fading}a, a 3 second portion of the captured mobile scenario is shown as viewed from the \gls{BS}. An arrow indicates the movement of one of the cars during this three second period. For one \gls{UE} in this car over the acquisition period shown, the channel magnitude of a single, vertically polarised \gls{BS} antenna was extracted, along with the respective diagonal element of the user side Gram matrix $\vec{H}_{r}^\mathsf{H}\vec{H}_{r}$ for one resource block $r$. Their magnitudes are plotted against each other in \figurename~\ref{fig_fading}b after normalization. It can be seen that the composite channel tends to follow the average of the single antenna case, smoothing out the faster fading extremities, and larger variations occur over the course of seconds rather than milliseconds.
\\
\begin{figure}[!t]
	\centering
	\includegraphics[width=\columnwidth]{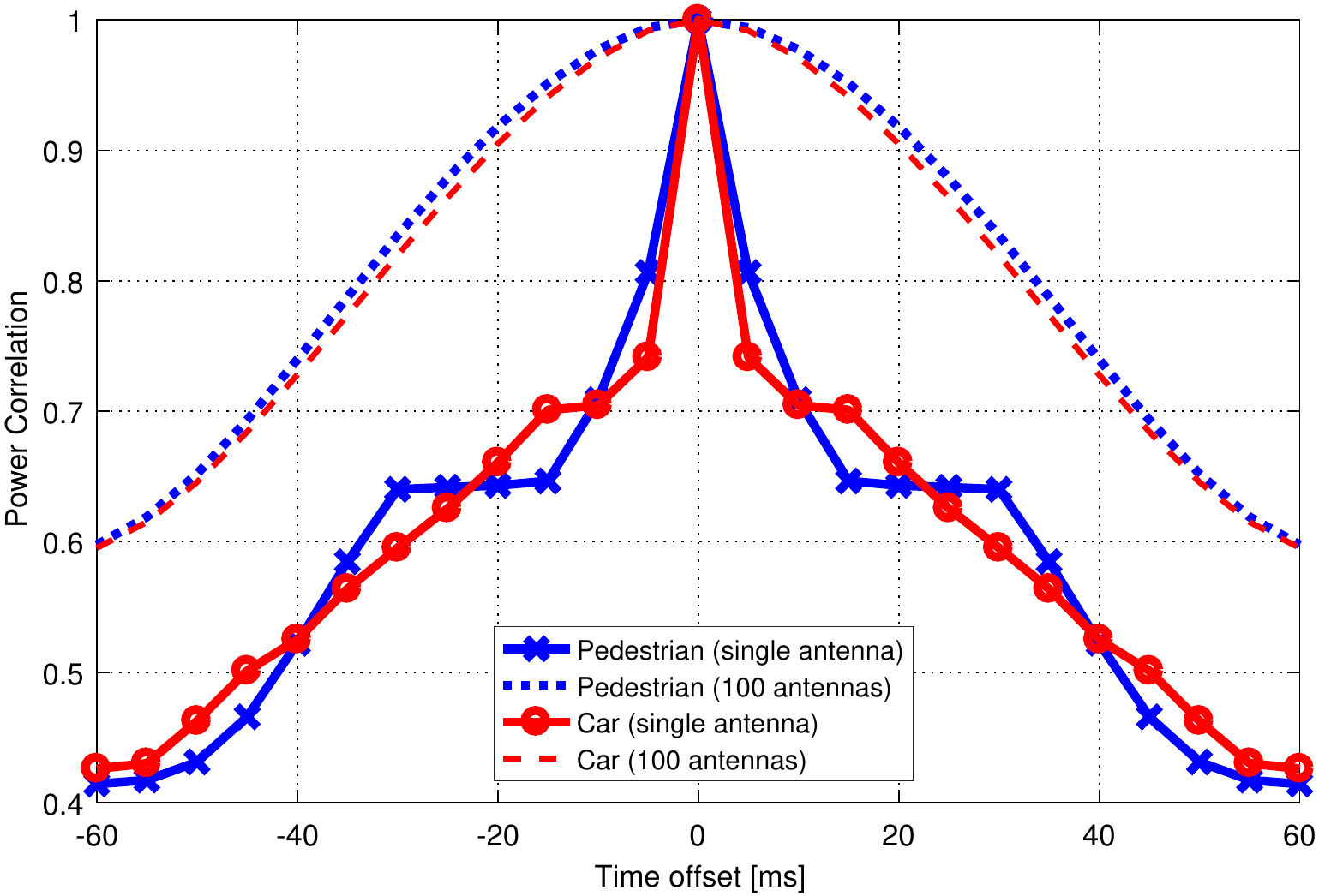}
	\caption{Correlation of the power for signal on one antenna versus 100 antennas for a pedestrian and car UE.}
	\label{fig_power_cor}
\end{figure}
\figurename~\ref{fig_power_cor} shows the correlation of the signal power over time offset for a pedestrian UE and a car UE when using either one antenna or 100 antennas.
Due to the channel hardening and the constructive combining of 100 signals, the power levels are much more stable as compared to the single antenna case. For this particular case, with 100 antennas, power control can be done at least five times slower than with a single antenna.
These two figures show, that performance of this nature not only demonstrates an improvement in robustness and latency due to the mitigation of fast-fade error bursts, but also that it is possible to greatly relax the update rate of power control when combining signals from many antennas.
\\
\subsubsection{Time Correlation}
\begin{figure}[!t]
	\centering
	\includegraphics[width=\columnwidth]{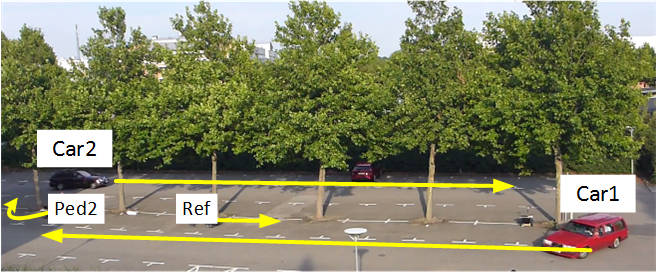}
	\caption{4 second subset of mobility scenario (\SI{10}{\second} to \SI{14}{\second} of full 30 second capture) used for temporal analysis. Arrows indicate the movement of each \gls{UE} over the \SI{4}{\second} duration. Car 2 does not exceed the maximum allowed speed of \SI{29}{\kilo\meter\per\hour}.}
	\label{fig_temp_scen}
\end{figure}
\begin{figure}[!t]
	\centering
	\includegraphics[width=\columnwidth]{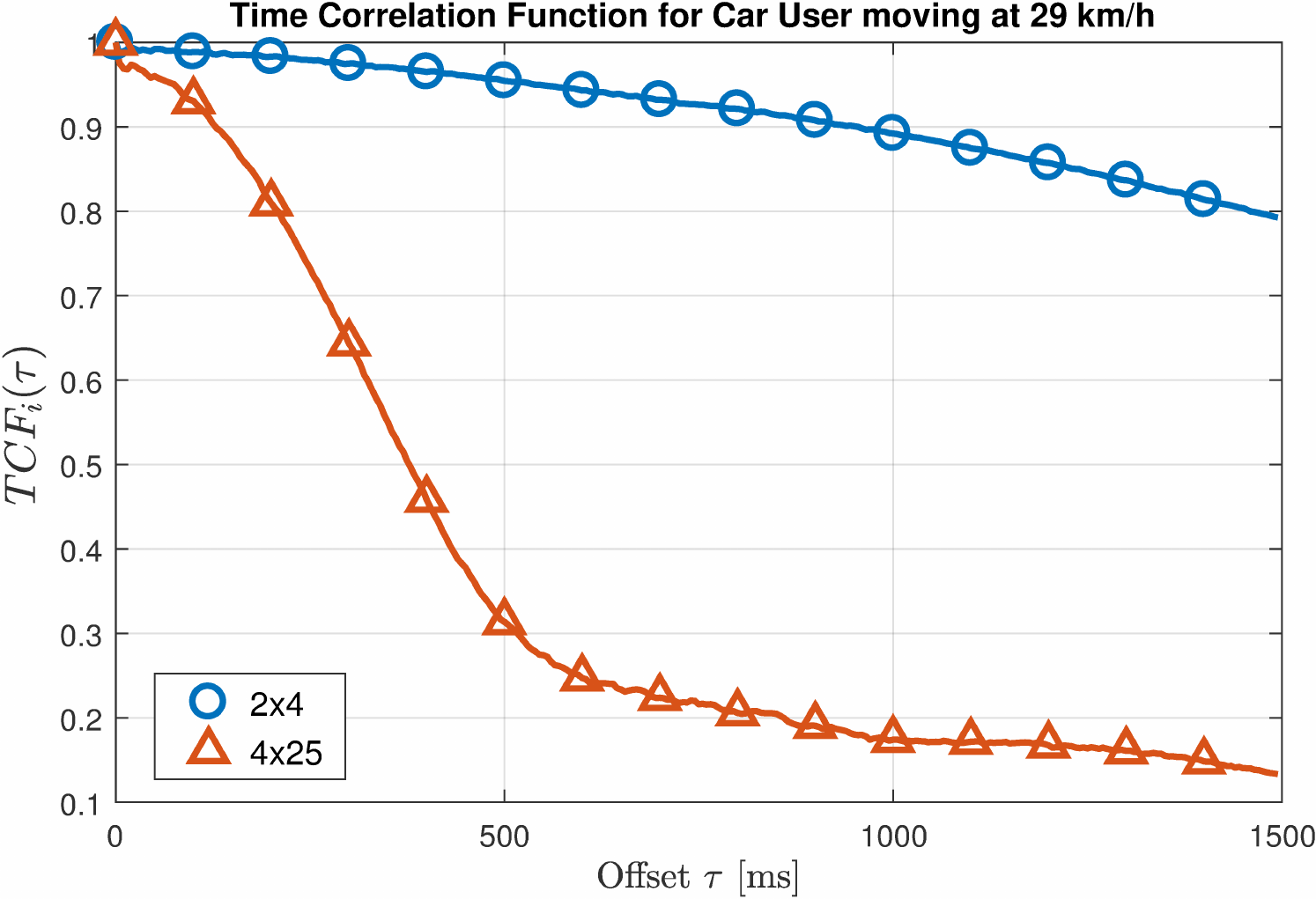}
	\caption{Correlation of the composite channel over time for all resource blocks of car 2 at a speed of \SI{29}{\kilo\meter\per\hour}. 100 antenna and 8 antenna cases are shown in 4x25 and 2x4 configurations respectively. Both curves are normalized to themselves, so array gain is not visible here.}
	\label{fig_TCF}
\end{figure}

As a \gls{UE} moves, it is of interest to view the correlation of the \gls{MIMO} channel vectors over time in order to ascertain how quickly the channel becomes significantly different. This will play a part in determining the required channel estimation periodicity for a given level of performance.
In \figurename~\ref{fig_temp_scen}, a \SI{4}{\second} second period of the 30 second mobility scenario is shown, with the arrows indicating the movement of each \gls{UE} during that period. Using one \gls{UE} from Car 2, the absolute values of the time correlation function for all resource blocks over the 4 second period are shown in \figurename~\ref{fig_TCF} for the first \SI{1.5}{\second} of movement.
Within the first \SI{500}{\milli\second} at this speed, the level of correlation has dropped significantly in the 100 antenna case to 0.3, whilst the 8 antenna case remains above 0.8 for the entire \SI{1.5}{\second} duration with a far shallower decay. For the 8 antenna and 100 antenna cases to become decorrelated by 20 percent, it takes \SI{1455}{\milli\second} and \SI{205}{\milli\second} respectively; a factor difference of approximately 7. This highlights how the precise spatial focusing of energy provided by massive \gls{MIMO} translates to a larger decorrelation in time from far smaller movements in space. The acceptable level of decorrelation will depend upon many factors such as the desired level of performance, the detection/precoding technique and the \gls{MCS}, but this result provides some insight into how rapidly a real channel vector can change in massive \gls{MIMO} under a moderate level of mobility.
\\
\subsubsection{Inter-User Interference}
\begin{figure}[!t]
	\centering
	\includegraphics[width=\columnwidth]{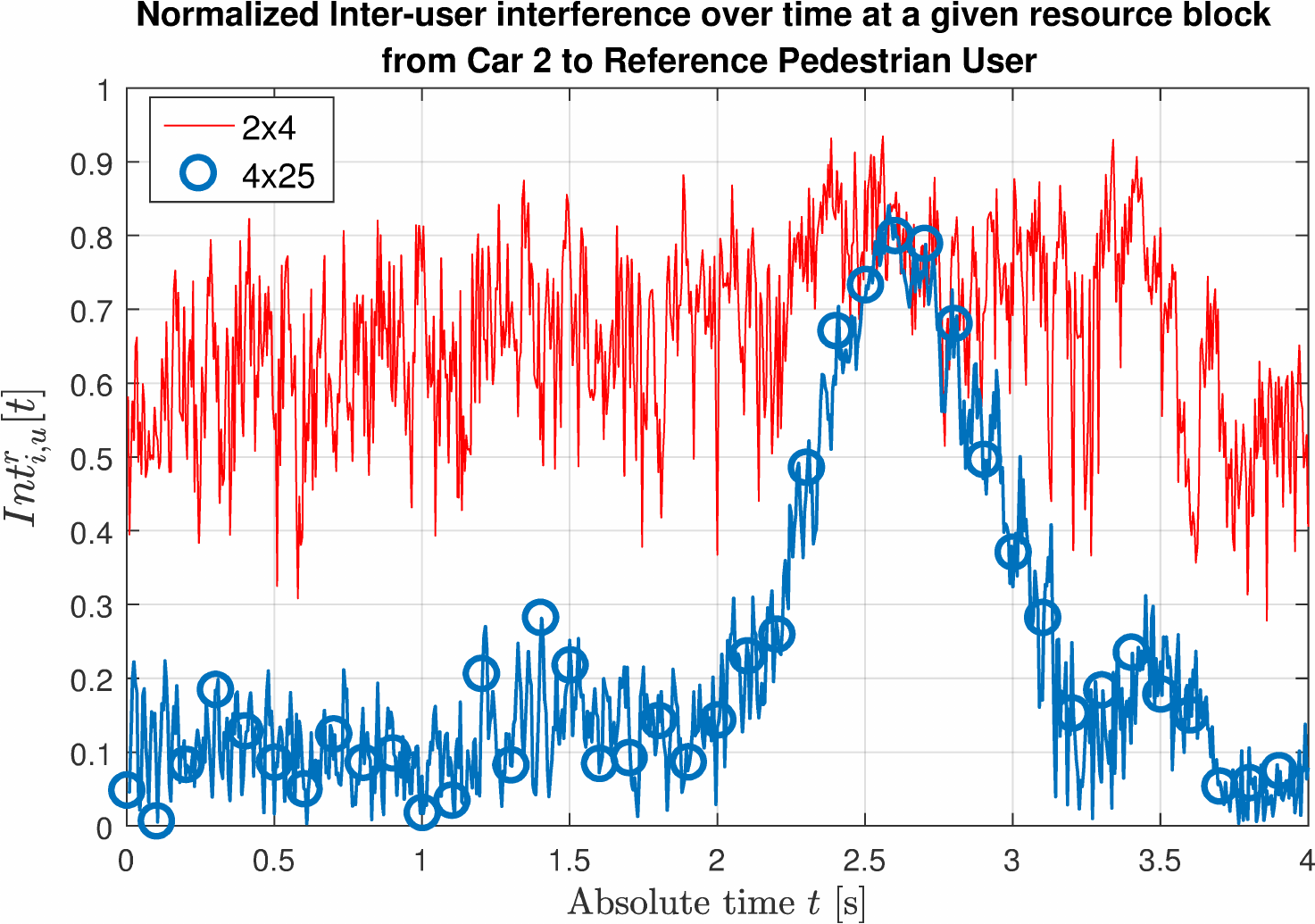}
	\caption{\gls{IUI} between car 2 and pedestrian user 1 over time}
	\label{fig_IUI_time}
\end{figure}

\begin{figure}[!t]
	\centering
	\includegraphics[width=\columnwidth]{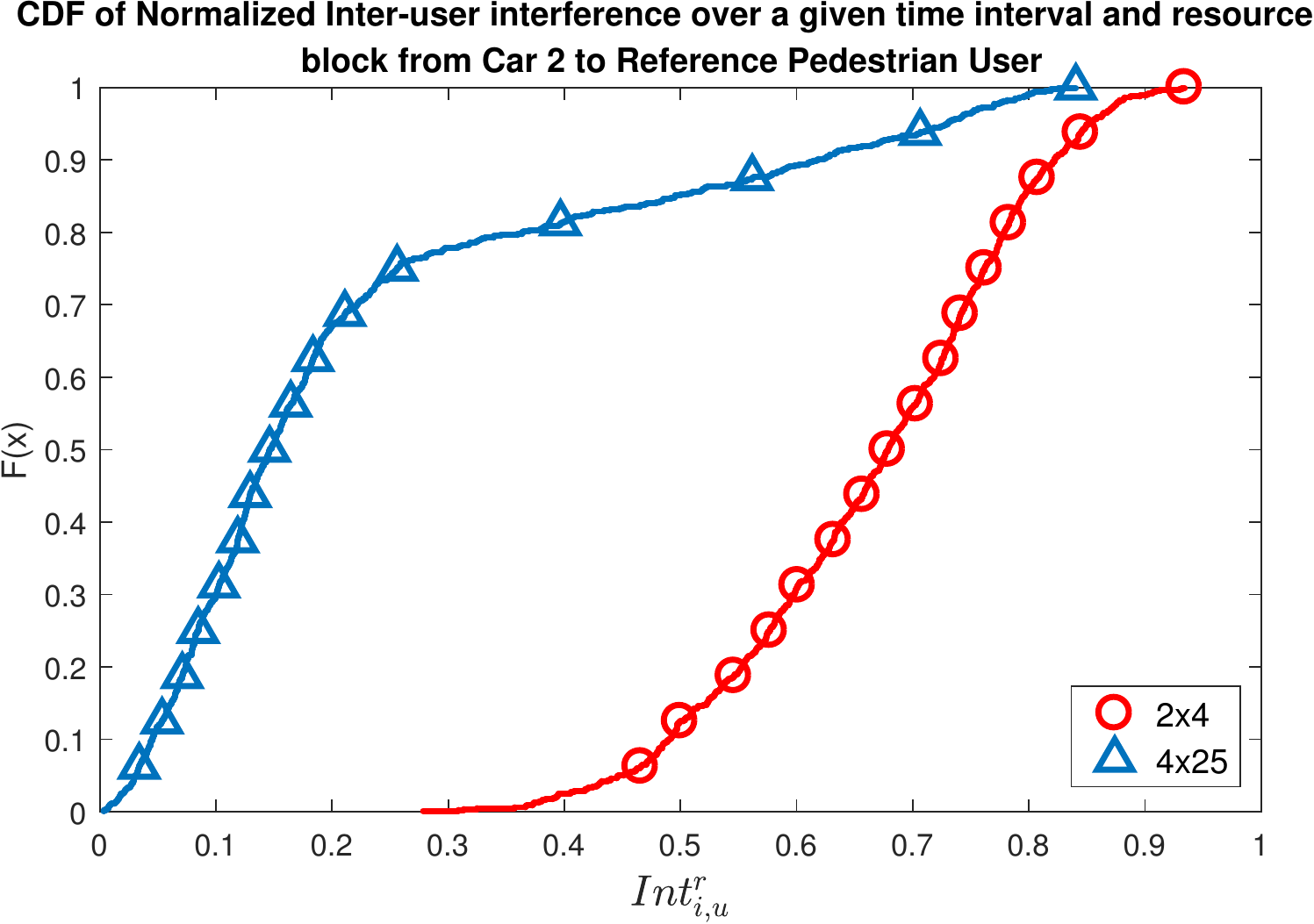}
	\caption{CDF of interference between car 2 and pedestrian user 1}
	\label{fig_IUI_CDF}
\end{figure}

Using the same \SI{4}{\second} scenario depicted in \figurename~\ref{fig_temp_scen}, the \gls{IUI} for the entire period was calculated between a single \gls{UE} in car 2 and a reference pedestrian, indicated as Ref in the figure. For clarity, it should be noted that \gls{IUI} here is a measure of user correlation; with \gls{ZF} precoding applied, the actual \gls{IUI} is small in practice. In \figurename~\ref{fig_IUI_time} the normalized \gls{IUI} is shown plotted over time for both 8 and 100 antennas, and in \figurename~\ref{fig_IUI_CDF} as an empirical \gls{CDF}.
In the 8 antenna case, the \gls{IUI} median is approximately 0.7 with a large variance from 0.28 up to 0.93. At \SI{2.5}{\second} into the scenario period, where it can be seen that car 2 will pass close to the reference user, the \gls{IUI} in \figurename~\ref{fig_IUI_time} does not appear to increase much beyond the average it is already maintaining. In the same graph for the 100 antenna case, the median \gls{IUI} is 0.15 and the level rises only significantly as the car passes close to the reference user. The level it peaks at is only a little under that of the 8 antenna case, but it rolls off to below 0.3 in approximately \SI{500}{\milli\second}. This interference anomaly can be seen in the \gls{CDF} of \figurename~\ref{fig_IUI_CDF} as a long upper tail, but the 75th percentile remains below 0.25. These results illustrate that about a two times reduction in the median level of \gls{IUI} between two users can be achieved in \gls{LOS} using 100 antennas over 8, but also that a smaller elevation resolution could limit the massive \gls{MIMO} benefits when users stack densely in a perpendicular line. However, it is likely that this could be mitigated further with intelligent user grouping.
\subsection{Uncoded BER Performance}
\begin{figure}[!t]
	\centering
	\includegraphics[width=\columnwidth]{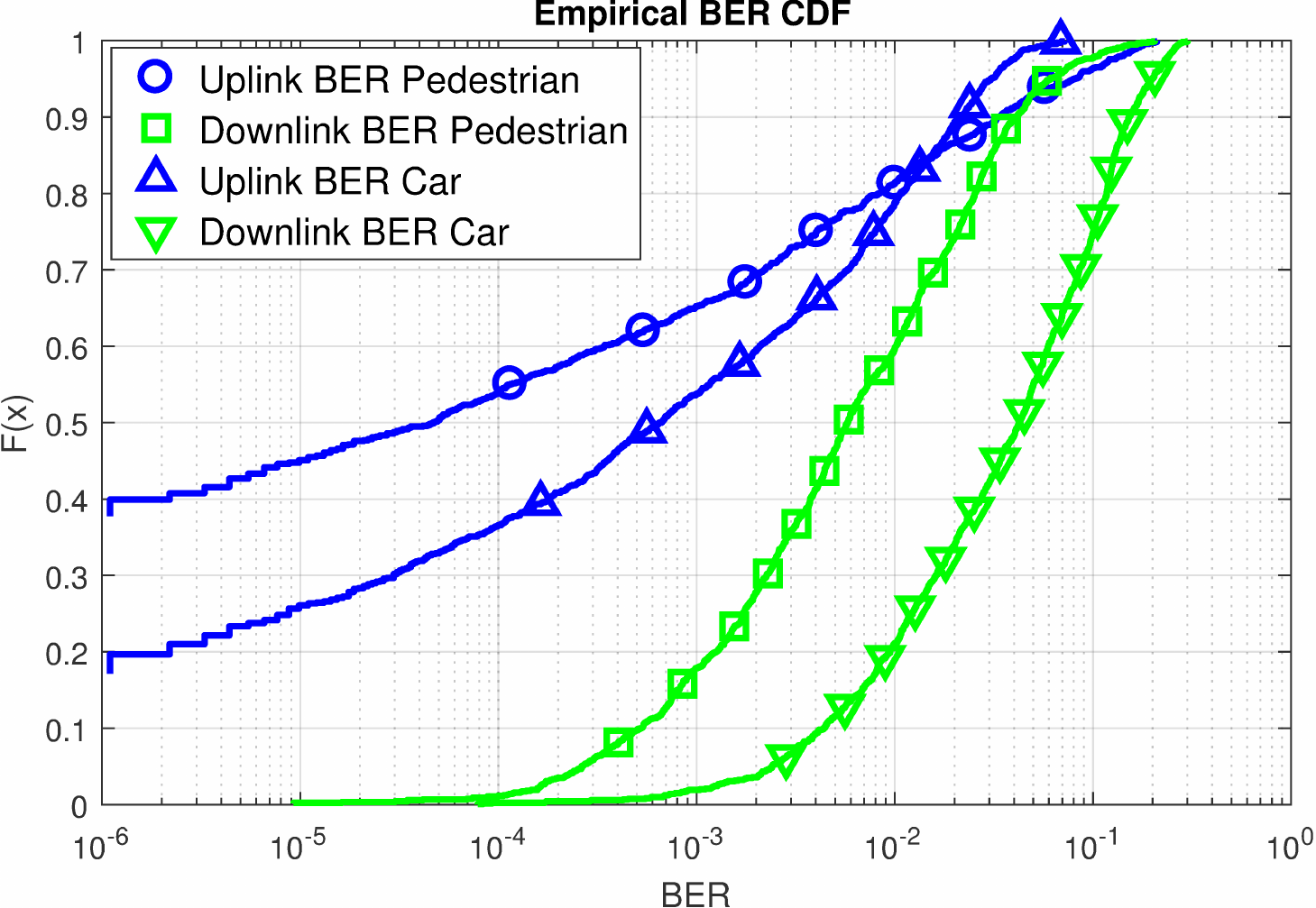}
	\caption{\gls{CDF} of uncoded \glspl{BER} for static and mobile scenarios using \gls{QPSK} and \gls{ZF}. 0.5 ms coherence interval, mobility up to \SI{50}{\kilo\meter\per\hour}.}
	\label{fig_BERs}
\end{figure}

In addition to the \gls{UL} pilot transmissions, the uncoded, real-time \glspl{BER} for both the \gls{UL} and \gls{DL} of each spatial mode using \gls{ZF} detection/precoding were recorded to provide an indication of the true system performance under mobility. As no power control or user grouping was applied, \gls{QPSK} modulation was used on each spatial stream for robustness. In ~\figurename~\ref{fig_BERs}, the \glspl{BER} for pedestrian and vehicular users in the mobile scenario are plotted as separate empirical \glspl{CDF} for both the \gls{UL} and \gls{DL}. The static scenario \glspl{BER} had a 95th percentile of 0 error, with the last 5 percent coming in below $10^{-5}$. This is stated here for reference purposes, but omitted from the \gls{CDF} plot. In first considering the \gls{UL}, it can be seen that the pedestrian \gls{BER} outperforms the vehicular up to the 80th percentile where both intersect at a \gls{BER} of $10^{-2}$, but the cars appear to suffer less in the upper extremities, tailing off early to a peak \gls{BER} of 7\%. This can be explained by the fact that the car antennas were roof mounted, well separated and clear of obstructions, whereas the pedestrian antenna pairs were closely spaced and occasionally shadowed by the person pushing the cart. Thus, whilst on average the higher level of mobility provided by the cars presents a greater challenge for detection, the worst performance is experienced by the pedestrian \glspl{UE} when their closely spaced \gls{USRP} antennas are simultaneously shadowed.
On the \gls{DL}, both curves are steeper and shifted further right than the \gls{UL} cases, which we would expect when both reciprocity calibration inaccuracies and channel ageing are considered. Looking at the pedestrian \gls{DL} case, the upper tail appears to slightly outperform the \gls{UL}. It is believed that this is because in the most extreme, shadowed cases, described above for the \gls{UL}, the transmit gain of the 100 \gls{BS} radios raises the terminal \gls{SNR} enough for this improvement.
This would also explain why, unlike the \gls{UL}, the \gls{DL} pedestrian \glspl{BER} do not outperform the vehicular in the upper 20th percentile. These differences aside, median \gls{DL} \glspl{BER} of 0.5\% and 4\% were observed for the pedestrian and vehicular cases respectively, and 90th percentiles of 4\% and 15\%. Whilst the absolute \gls{SNR} at each \gls{UE} or \gls{BS} antenna was not measured, these results still illustrate that the system was able to track the channel accurately enough to maintain 8 reliable spatial streams using under moderate mobility with no user grouping or power control. With the latter enhancements in place and an error-correcting code, it is highly likely that the system could provide satisfying performance, even with higher-order modulation schemes.

\section{Conclusions}
\label{conc}
The performance of a $100\times8$ real-time massive \gls{MIMO} system operating in \gls{LOS} with moderate mobility has been presented and analysed. To the best of the authors' knowledge, these are the first results of their kind that begin to indicate the performance of massive \gls{MIMO} as the composite channel changes over the course of a more dynamic scenario. The \gls{SVS} results indicate that azimuth resolution is still to be preferred overall in a \gls{LOS} situation with minimal scatters present, but the difference is perhaps not significant enough to warrant the inherent cost and complexity involved when deploying a physically long array. The \gls{IUI} results also indicate that lack of elevation resolution could cause problems for users densely bunched in a perpendicular line, but situations like this may be improved with intelligent user grouping. When considering the correlation of mobile channel vectors over time, it was shown that the 100 antenna case decorrelated by 20\% 7 times faster than the 8 antenna case, providing an indication of the spatial focusing effect in a \gls{MIMO} system. Finally, for the mobile scenario illustrated, it was shown that \glspl{BER} of 1\% and 10\% for the 80th percentile can be achieved for \gls{UL} and \gls{DL} respectively, using \gls{QPSK} with \gls{ZF} and no form of power control, user grouping or error-correcting code.


%

\appendices


\section*{Acknowledgment}
The authors wish to thank all academic staff and post graduate students involved who contributed to the measurement trial operations. They also acknowledge the financial support of the \gls{EPSRC} (EP/I028153/1), the European Union Seventh Framework Programme (FP7/2007-2013) under grant agreement no. 619086 (MAMMOET), NEC, NI, the Swedish Foundation for Strategic Research and the Strategic Research Area ELLIIT.

\ifCLASSOPTIONcaptionsoff
  \newpage
\fi



\bibliographystyle{MyIEEEtran}
\bibliography{library}
%
%
%
\begin{IEEEbiography}
	[{\includegraphics[width=1in,height=1.25in,clip,keepaspectratio]{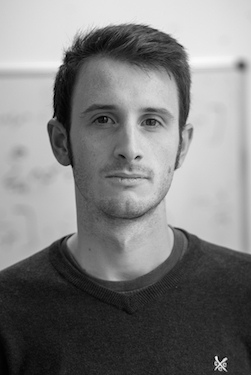}}]{Paul Harris}
received the B.Eng. degree in Electronic Engineering from the University of Portsmouth in 2013 and joined the Communication Systems \& Networks Group at the University of Bristol in the same year to commence a PhD. His research interests include massive \gls{MIMO} system design, real-world performance evaluation and the optimisation of techniques or algorithms using empirical data. Working in collaboration with Lund University and National Instruments, he implemented a 128-antenna massive MIMO test system and led two research teams to set spectral efficiency world records in 2016. For this achievement, he received 7 international awards from National Instruments, Xilinx and Hewlett Packard Enterprise.
\end{IEEEbiography}
\begin{IEEEbiography}[{\includegraphics[width=1in,height=1.25in,clip,keepaspectratio]{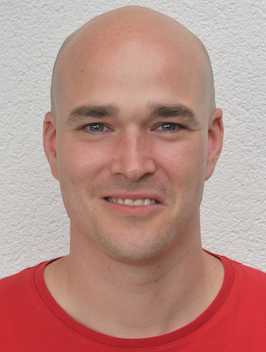}}]{Steffen Malkowsky}
	received the B.Eng. degree in Electrical Engineering and Information Technology from Pforzheim University, Germany in 2011 and the M.Sc. degree in Electronic Design from Lund University in 2013. He is currently a PhD student in the Digital ASIC group at the department of Electrical and Information Technology, Lund University. His research interests include development of reconfigurable hardware and implementation of algorithms for wireless communication with emphasis on massive MIMO. For the development of a massive MIMO testbed in collaboration with the University of Bristol and National Instruments, and a spectral efficiency world record, he received 7 international awards from National Instruments, Xilinx and Hewlett Packard Enterprise.
\end{IEEEbiography}
\begin{IEEEbiography}[{\includegraphics[width=1in,height=1.25in,clip,keepaspectratio]{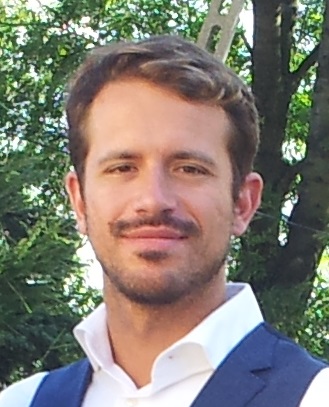}}]{Joao Vieira}
	received the B.Sc. degree in Electronics and Telecommunications Engineering from University of Madeira in 2011, and the M.Sc. degree in Wireless Communications from Lund University, Sweden in 2013. He is currently working towards a Ph.D. degree at the department of Electrical and Information Technology in Lund University. His main research interests regard parameter estimation and implementation issues in massive MIMO systems.
\end{IEEEbiography}
\begin{IEEEbiography}[{\includegraphics[width=1in,height=1.25in,clip,keepaspectratio]{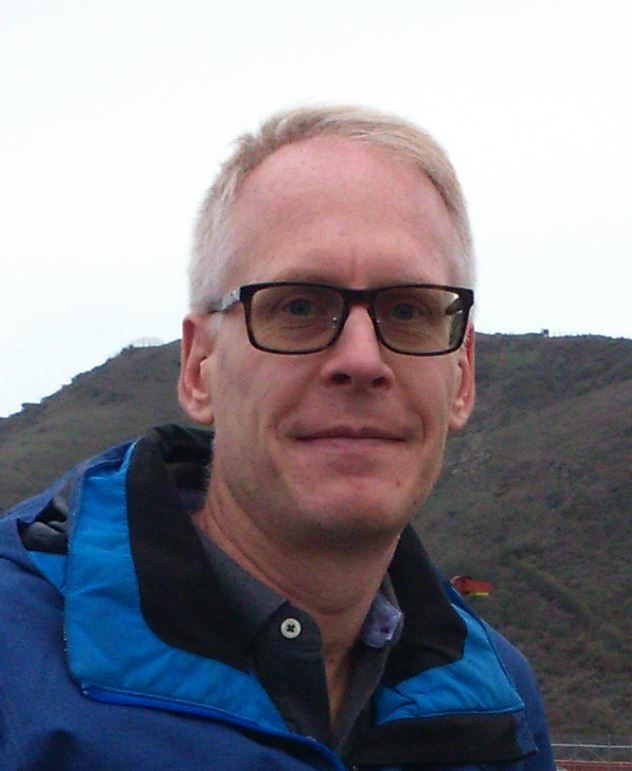}}]{Erik Bengtsson}
	received M.Sc. in Electrical Engineering from Lund University 1997. He joined Ericsson Mobile Communication AB in Lund the same year and worked with RF ASIC design until 2005. He then joined Nokia A/S in Copenhagen and worked with antenna concept development with focus on reconfigurable antennas. In 2011 he joined Sony Mobile in Lund and belongs to the Network Technology lab. From 2015 he is an industry PhD student at the Department of Electrical and Information Technology, Lund University, partly founded by Swedish Foundation for Strategic Research (SSF). His current research focus is terminal diversity aspects from a massive MIMO perspective.
\end{IEEEbiography}
\begin{IEEEbiography}[{\includegraphics[width=1in,height=1.25in,clip,keepaspectratio]{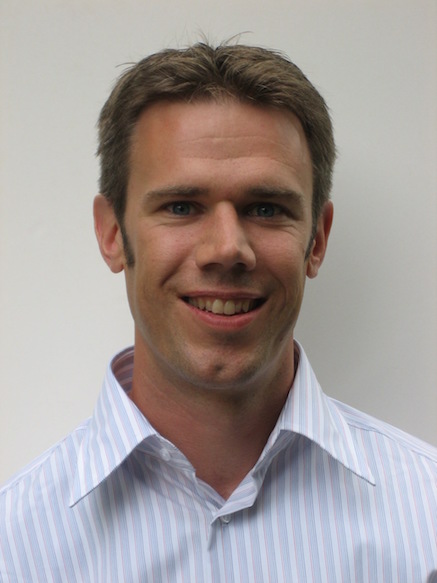}}]{Fredrik Tufvesson}
	received his Ph.D. in 2000 from Lund University in Sweden. After two years at a startup company, he joined the department of Electrical and Information Technology at Lund University, where he is now professor of radio systems. His main research interests are channel modelling, measurements and characterization for wireless communication, with applications in various areas such as massive MIMO, UWB, mm wave communication, distributed antenna systems, radio based positioning and vehicular communication. Fredrik has authored around 60 journal papers and 120 conference papers, recently he got the Neal Shepherd Memorial Award for the best propagation paper in IEEE Transactions on Vehicular Technology.
\end{IEEEbiography}
\begin{IEEEbiography}
	[{\includegraphics[width=1in,height=1.25in,clip,keepaspectratio]{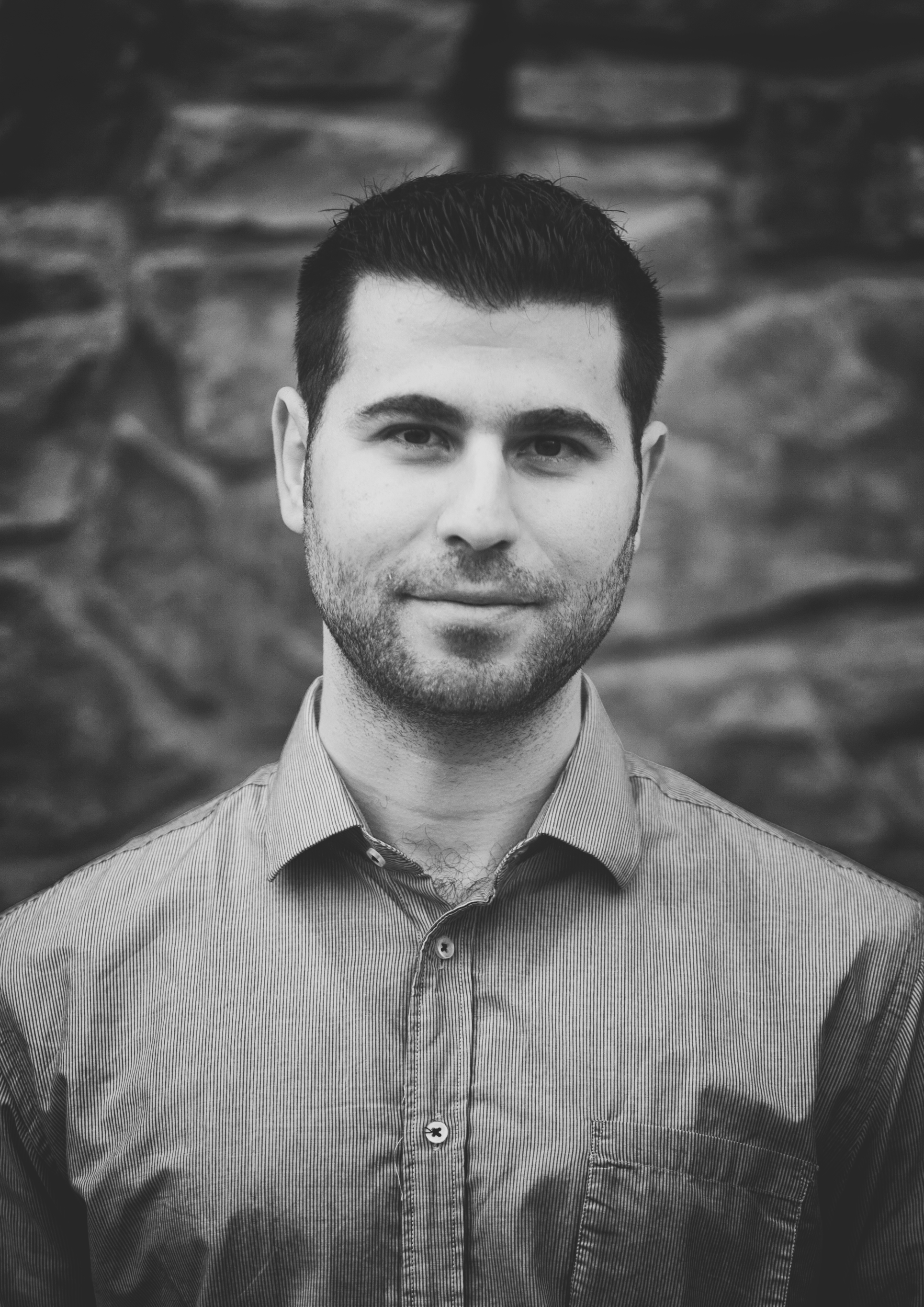}}]{Wael Boukley Hasan}
	received his M.Sc. in Mobile Communications Engineering with distinction from Heriot-Watt University in 2013. After graduating, he worked for one year in Alcatel-Lucent in the Small Cells Platform Development Department. He then joined the CDT in Communications at the University of Bristol in 2014. His research within the CSN group is focused on investigating and developing different techniques for massive \gls{MIMO}, with an interest in increasing spectral efficiency and power efficiency. He was a member of the University of Bristol research team that set spectral efficiency world records in 2016 in collaboration with Lund University.
\end{IEEEbiography}
\begin{IEEEbiography}[{\includegraphics[width=1in,height=1.25in,clip,keepaspectratio]{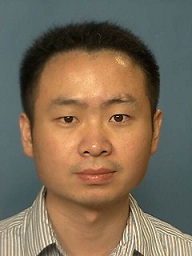}}]{Liang Liu}
	received his B.S. and Ph.D. degree in the Department of Electronics Engineering (2005) and Micro-electronics (2010) from Fudan University, China. In 2010, he was with Rensselaer Polytechnic Institute (RPI), USA as a visiting researcher. He joined Lund University as a Post-doc in 2010. Since 2016, he is Associate Professor at Lund University. His research interest includes wireless communication system and digital integrated circuits design. He is a board member of the IEEE Swedish SSC/CAS chapter. He is also a member of the technical committees of VLSI systems and applications and CAS for communications of the IEEE circuit and systems society.
\end{IEEEbiography}
\begin{IEEEbiography}
	[{\includegraphics[width=1in,height=1.25in,clip,keepaspectratio]{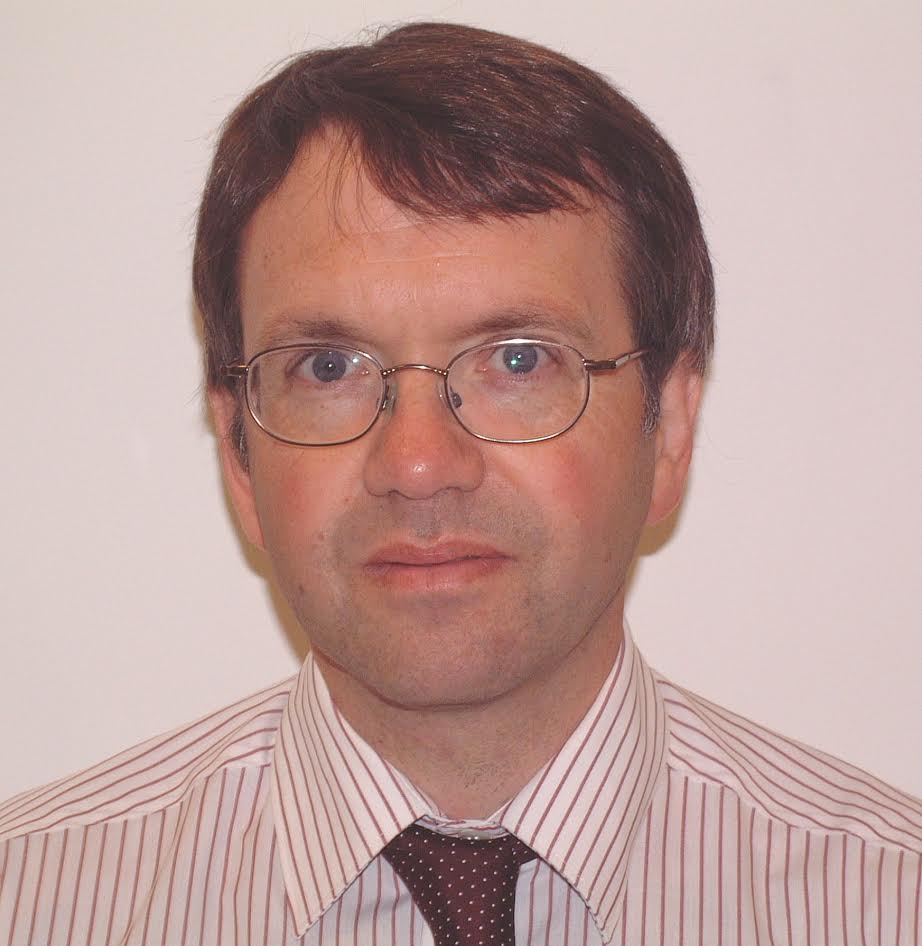}}]{Mark Beach}
received his PhD for research addressing the application of Smart Antenna techniques to GPS from the University of Bristol in 1989, where he subsequently joined as a member of academic staff. He was promoted to Senior Lecturer in 1996, Reader in 1998 and Professor in 2003. He was Head of the Department of Electrical \& Electronic Engineering from 2006 to 2010, and then spearheaded Bristol's hosting of the EPSRC Centre for Doctoral Training (CDT) in Communications. He currently manages the delivery of the CDT in Communications, leads research in the field of enabling technologies for the delivery of 5G and beyond wireless connectivity, as well as his role as the School Research Impact Director. Mark's current research activities are delivered through the Communication Systems and Networks Group, forming a key component within Bristol's Smart Internet Lab. He has over 25 years of physical layer wireless research embracing the application of Spread Spectrum technology for cellular systems, adaptive or smart antenna for capacity and range extension in wireless networks, MIMO aided connectivity for through-put enhancement, Millimetre Wave technologies as well as flexible RF technologies for SDR modems underpins his current research portfolio.
\end{IEEEbiography}
\begin{IEEEbiography}[{\includegraphics[width=1in,height=1.25in,clip,keepaspectratio]{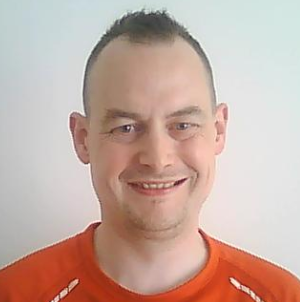}}]{Simon Armour}
Simon Armour received the B.Eng. degree in Electronics and Communication Engineering from the University of Bath, Bath, U.K., in 1996, and the Ph.D. degree in Electrical and Electronic Engineering from the University of Bristol, Bristol, U.K., in 2001. Since 2001, he has been a Member of Academic Staff with the University of Bristol, where, since 2007, he has been a Senior Lecturer. He has authored or co-authored over 100 papers in the field of baseband Pysical layer and Medium Access Control layer techniques for wireless communications with a particular focus on \gls{OFDM}, coding, \gls{MIMO}, and cross-layer multiuser radio resource management strategies. He has investigated such techniques in general terms, as well as specific applications to Wireless Local Area Networks and cellular networks.
\end{IEEEbiography}
\begin{IEEEbiography}[{\includegraphics[width=1in,height=1.25in,clip,keepaspectratio]{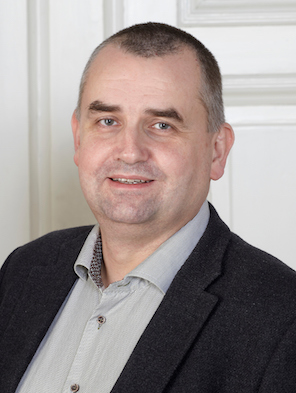}}]{Ove Edfors}
	is Professor of Radio Systems at the Department of Electrical and Information Technology, Lund University, Sweden.
	His research interests include statistical signal processing and low-complexity algorithms with applications in wireless communications. 
	In the context of Massive MIMO, his main research focus is on how realistic propagation characteristics 
	influence system performance and base-band processing complexity.
\end{IEEEbiography}






\end{document}